\documentclass[11pt]{article}

\usepackage{times}
\usepackage{color,graphicx}
\usepackage{enumerate}
\usepackage{epsfig,subfigure,setspace}
\usepackage{multirow}
\usepackage{ulem,float}
\usepackage{amsmath,amssymb,latexsym,url,amsthm}
\usepackage{harvard}


\newtheorem{myalg}{Algorithm}

\usepackage[explicit]{titlesec}
\usepackage{lipsum}

\definecolor{subtler}{rgb}{1,0,0.1}    

\newcommand{\be}{\begin{equation}}
\newcommand{\ee}{\end{equation}}
\newcommand{\ba}{\begin{eqnarray}}
\newcommand{\ea}{\end{eqnarray}}

\newcommand{\ban}{\begin{eqnarray*}}
\newcommand{\ean}{\end{eqnarray*}}
\newcommand{\ben}{\begin{equation*}}
\newcommand{\een}{\end{equation*}}
\def\em{\textit}
\def\rms{\smallskip\noindent\textit} 

\def\mt{{\mathcal {T}}}
\def\mp{{\mathcal P}} 

\def\mo{{\mathcal O}}
\def\mt{{\mathcal T}}
\def\ra{{\rm A}}
\def\rc{{\rm C}}
\def\rb{{\rm B}}
\def\rg{{\rm G}}
\def\rd{{\rm D}}
\def\rf{{\rm F}}
\def\re{{\rm E}}
\def\ret{{\rm E_T}}
\def\mc{\mathcal{C}} 
\def\ms{\mathcal{S}} 
\def\ds{{\rm DS}}
\def\pT{p_{\rm T}}
\def\pb{p_{\rm b}}
\def\scats{SCATS}
\def\nt{$\mathbb{NT}$} 
\def\tu{$\mathbb{PU}$} 
\def\tc{$\mathbb{PC}$} 
\def\au{$\mathbb{AU}$} 
\def\ac{$\mathbb{AC}$} 

\graphicspath{
{Figures/}
}

\def\threeup{0.325}

\def\phases										{\includegraphics[scale=0.13]{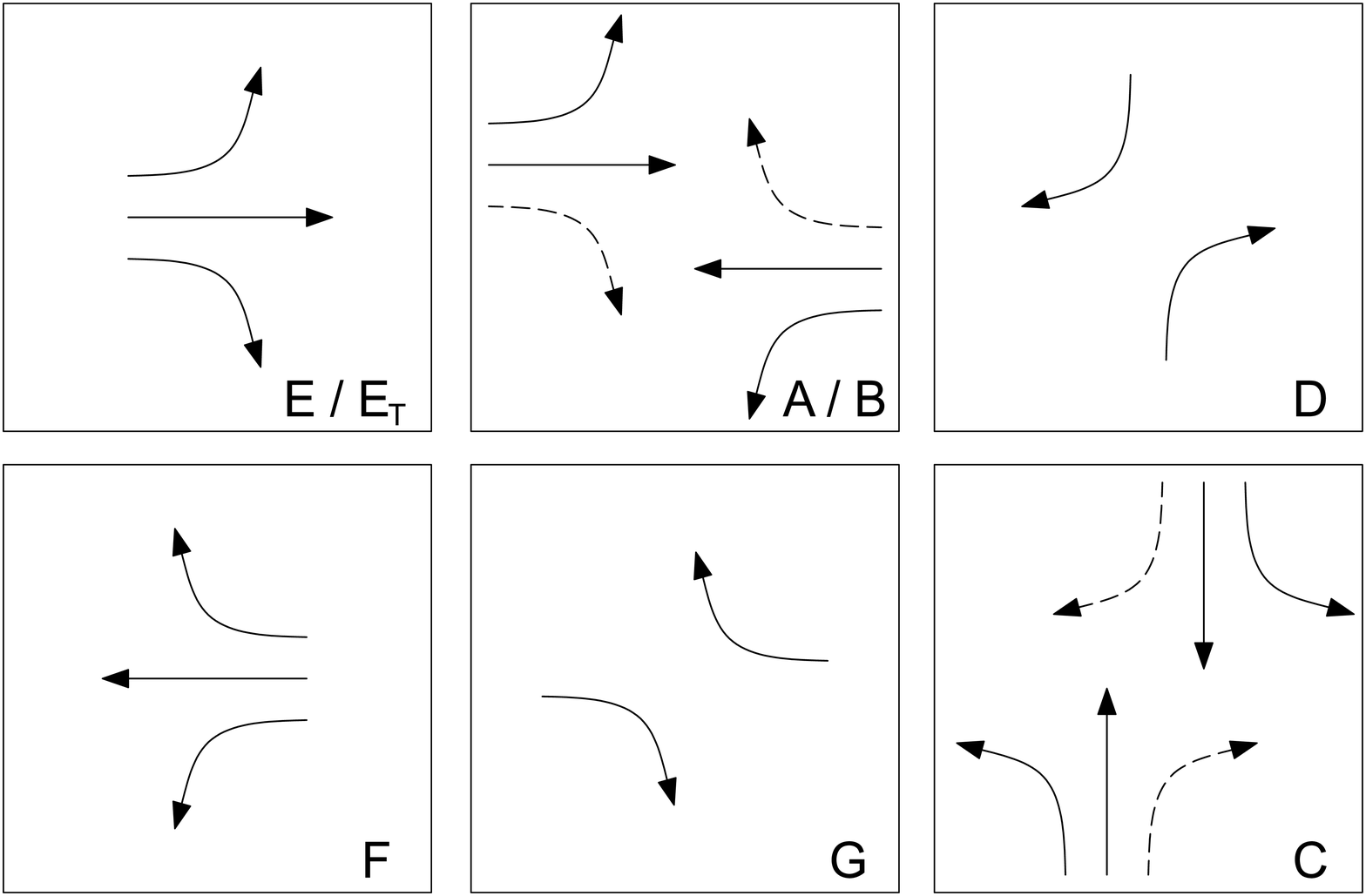}}
\def\network									{\includegraphics[scale=0.25]{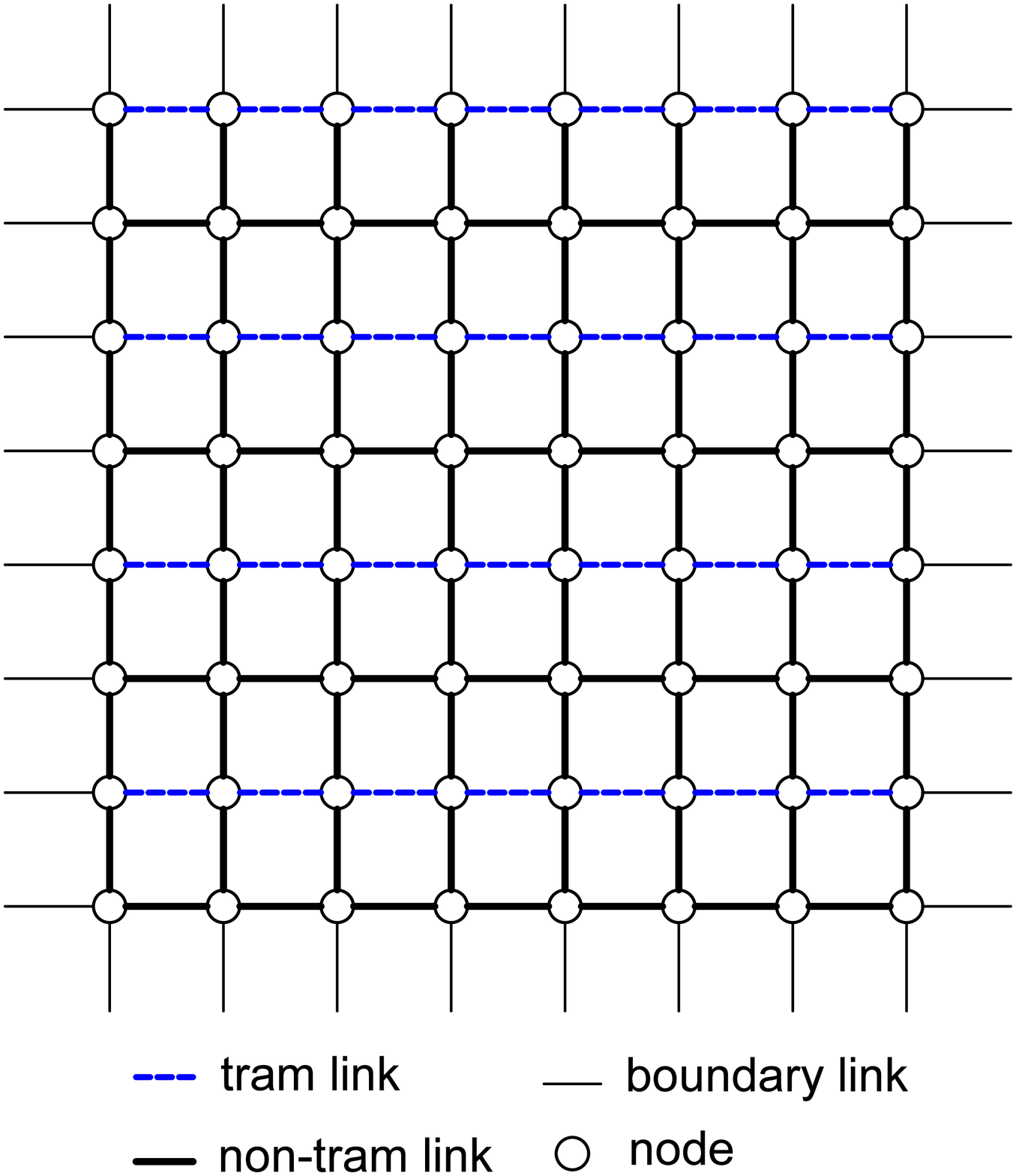}}
\def\tramIntersection							{\includegraphics[scale=0.35]{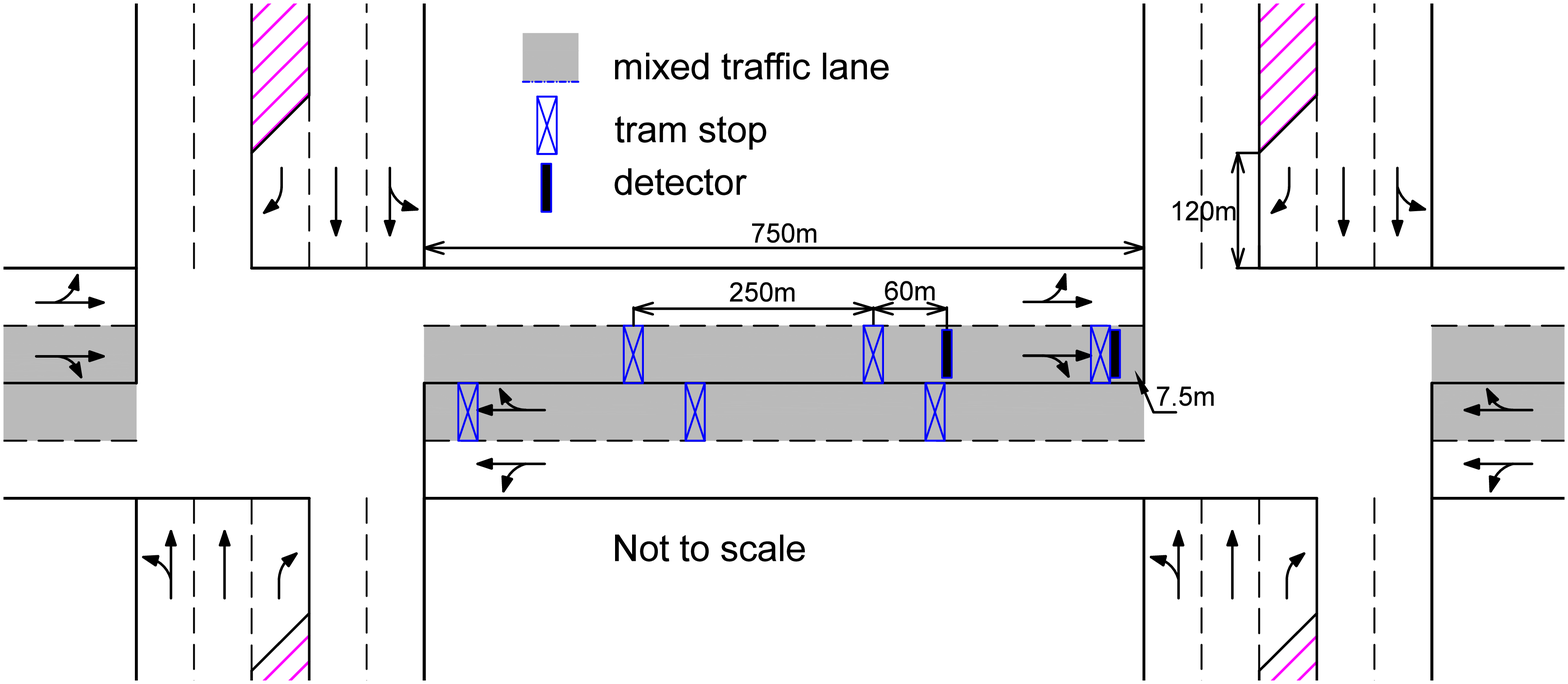}}

\def\linking									{\includegraphics[scale=0.25]{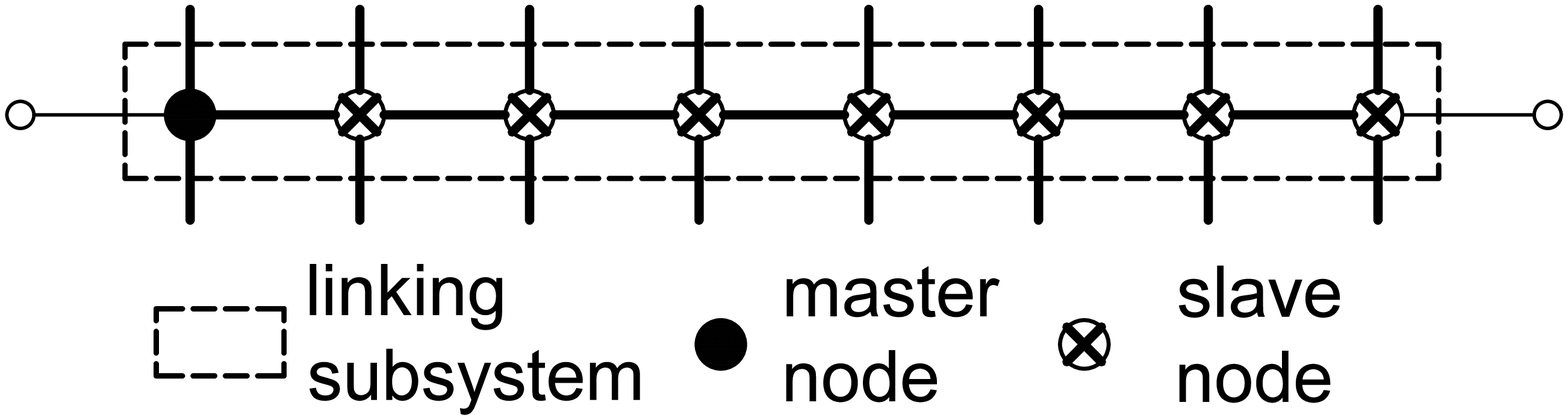}}

\def\googlemap									{\includegraphics[scale=0.26]{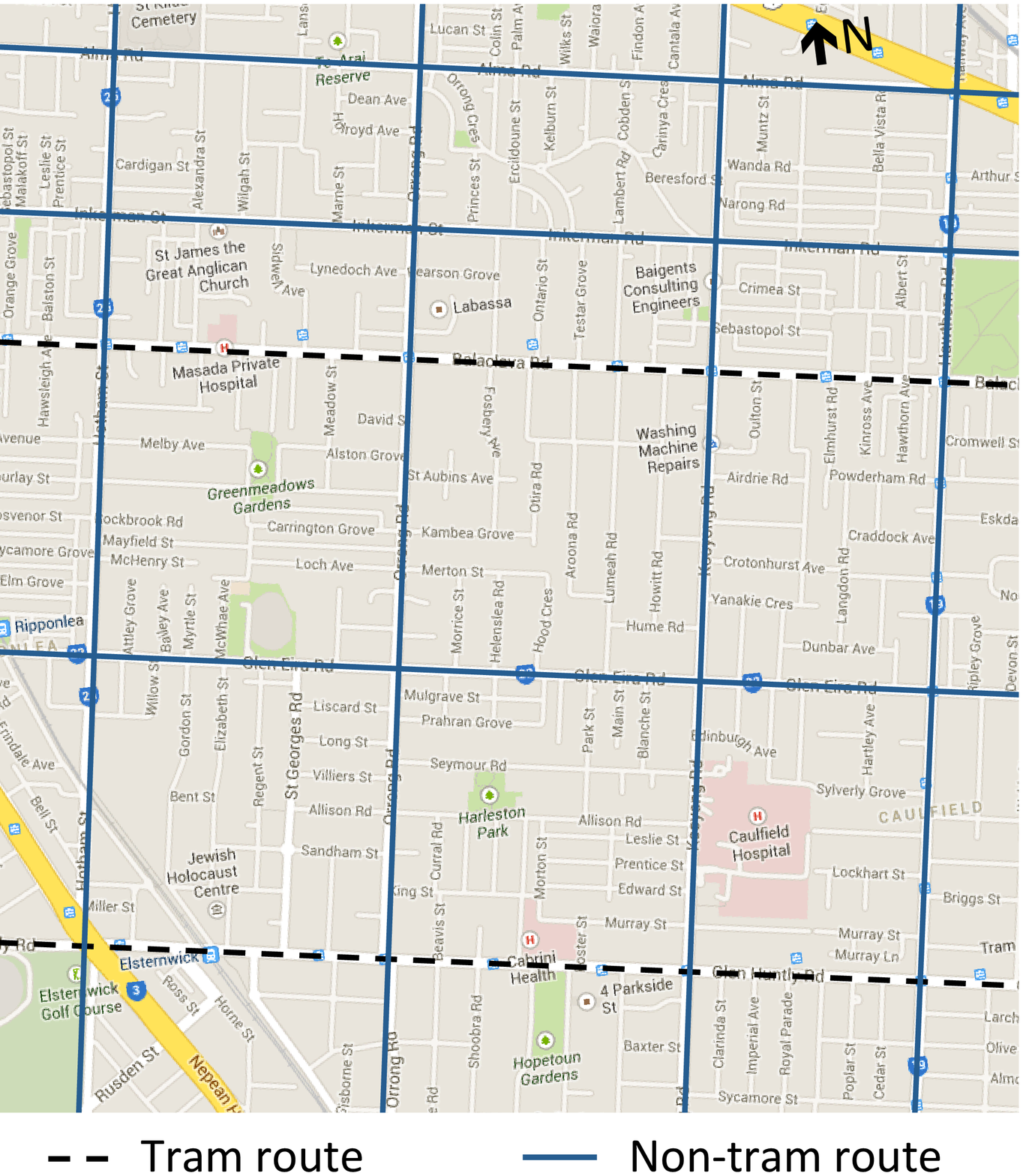}}

\def\densityprofileUS						{\includegraphics[scale=\threeup]{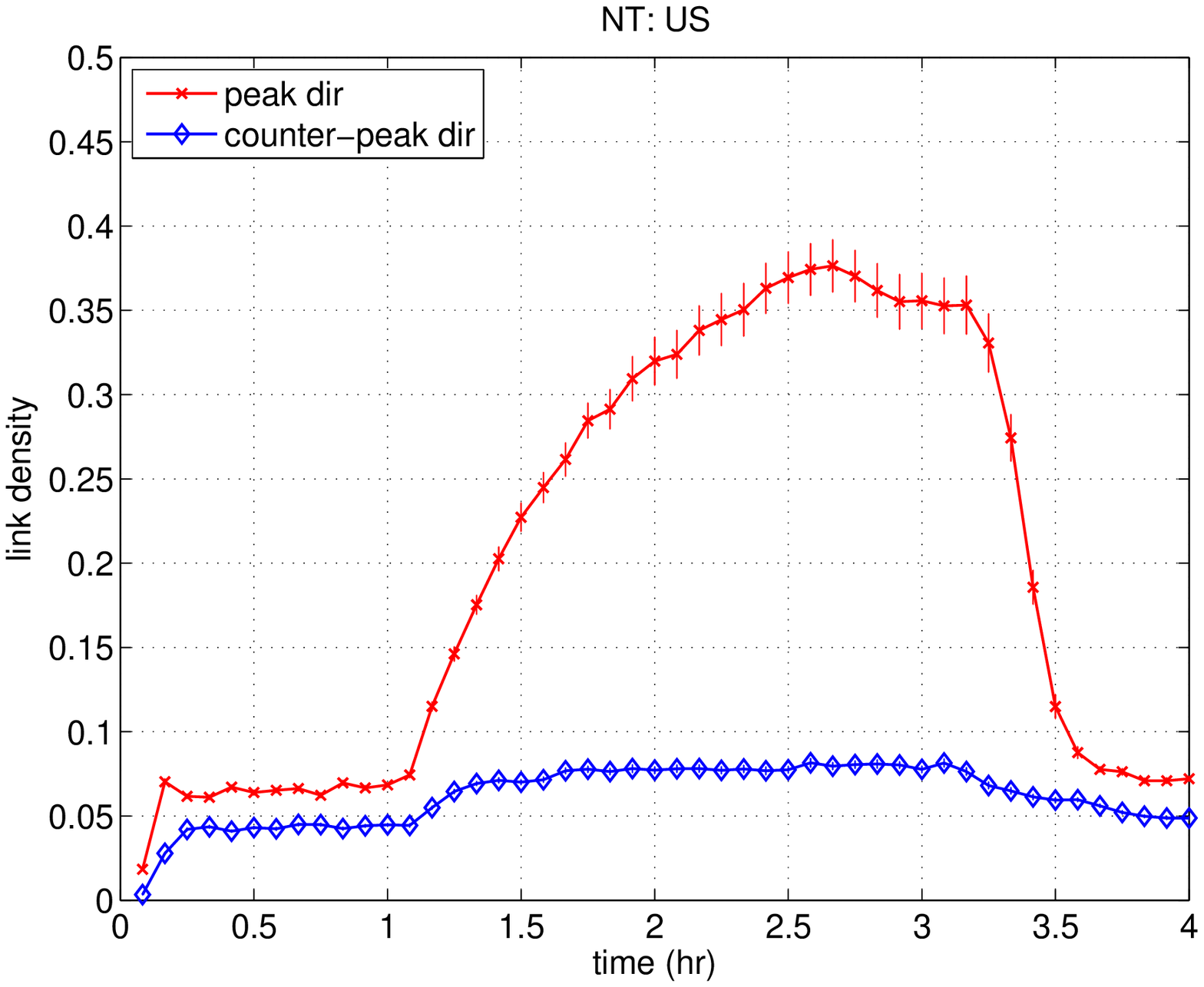}}
\def\densityprofileOS						{\includegraphics[scale=\threeup]{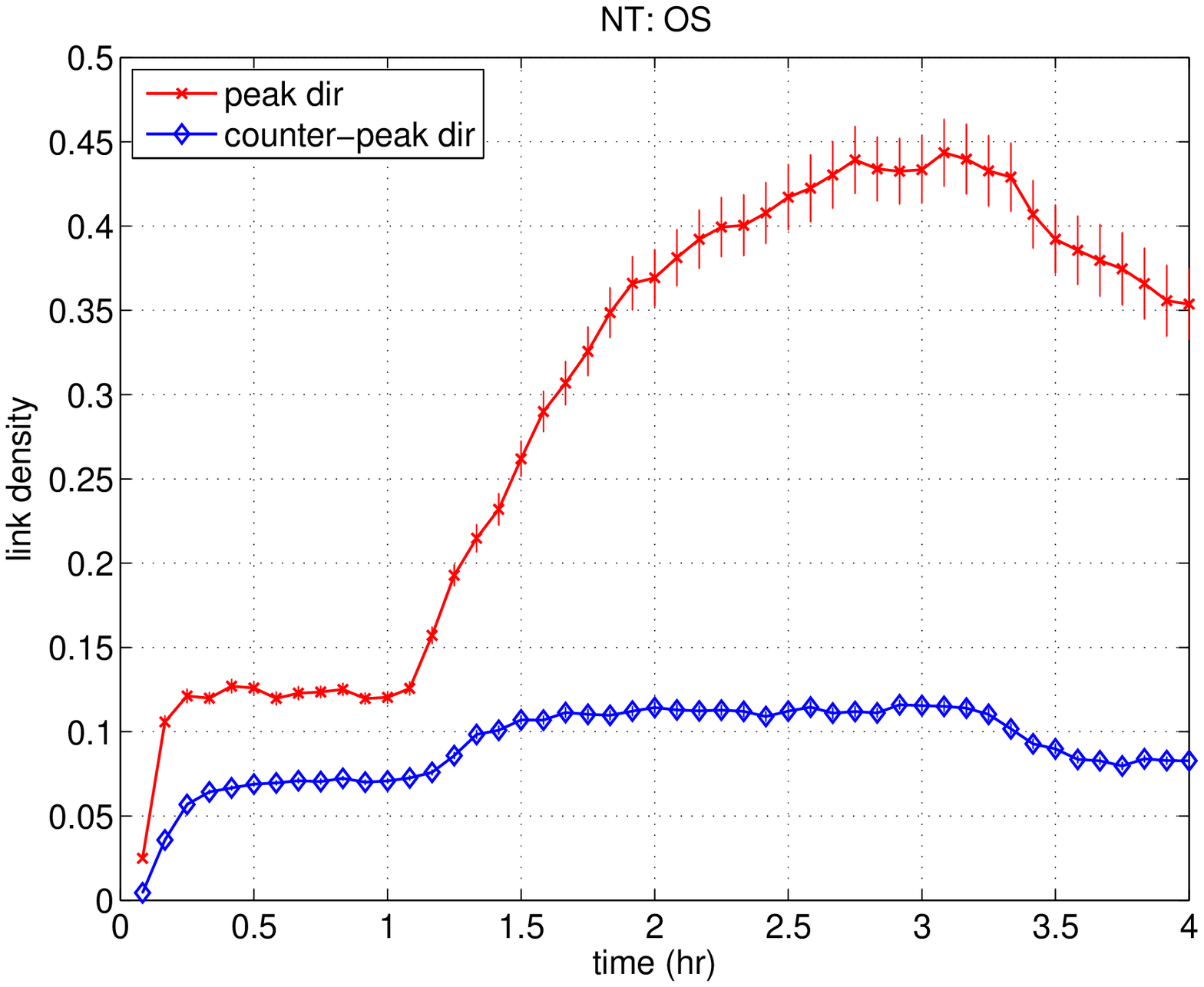}}

\def\tramageOS								{\includegraphics[scale=\threeup]{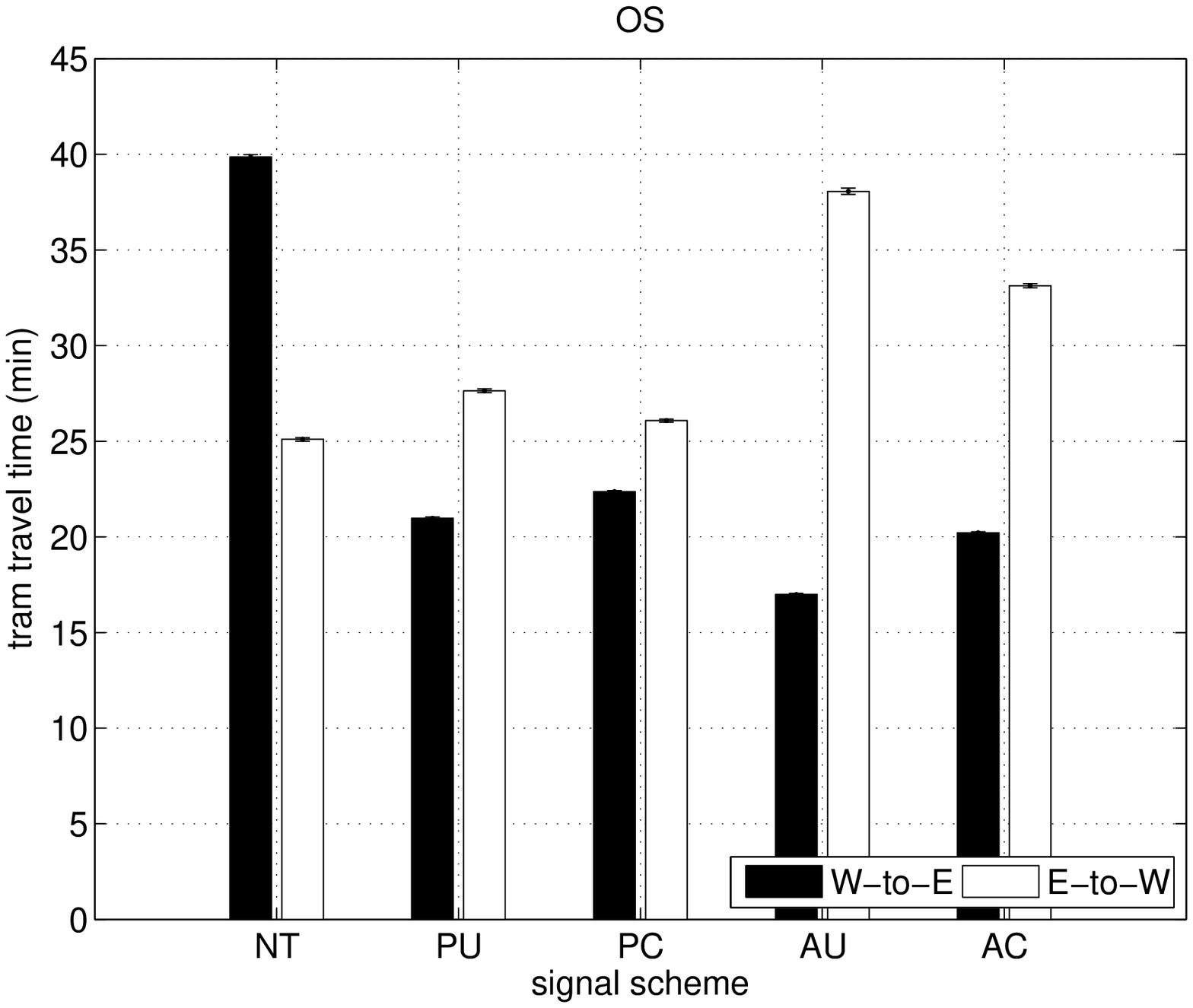}}
\def\tramageUS								{\includegraphics[scale=\threeup]{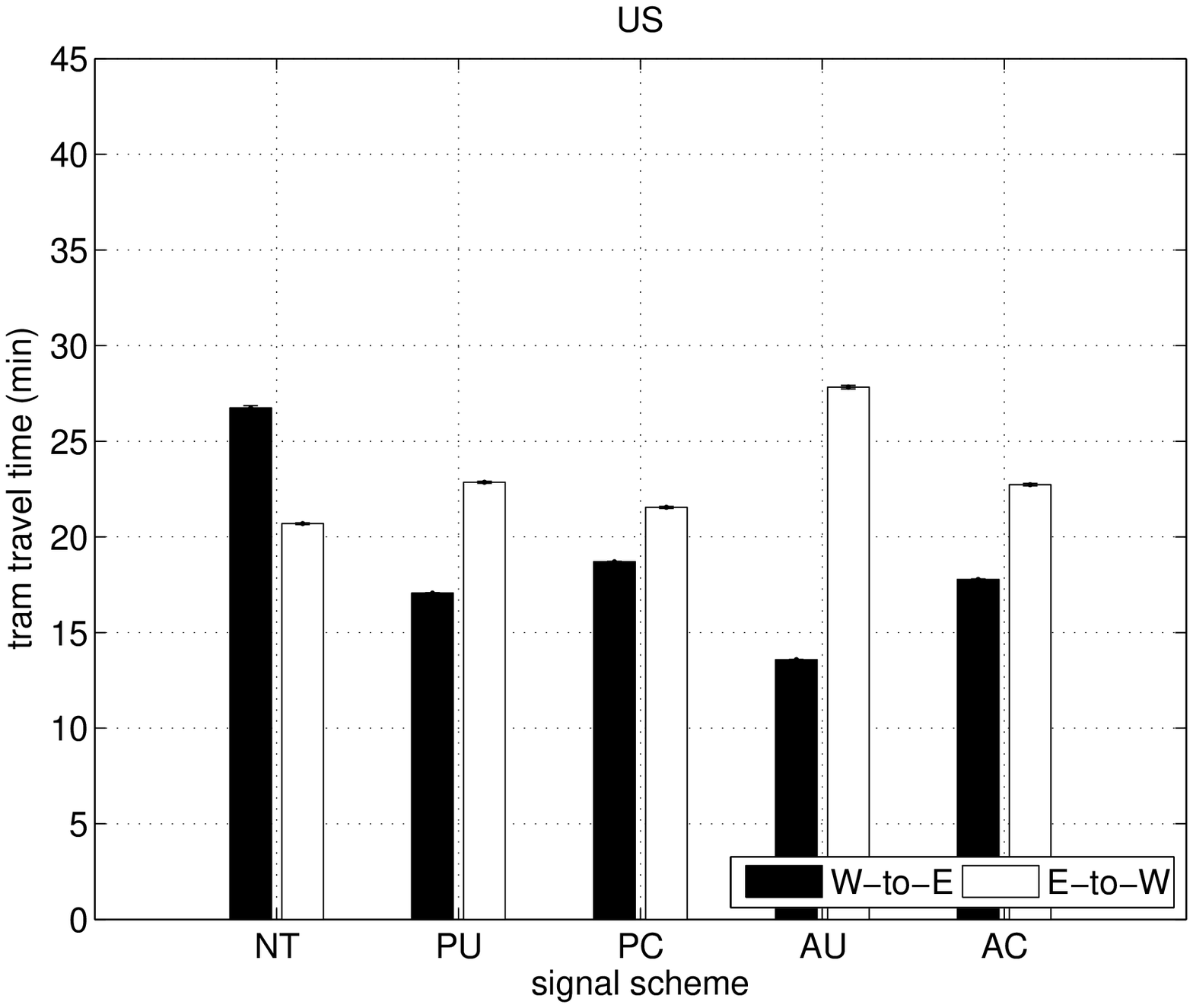}}
\def\tramthroughputOS						{\includegraphics[scale=\threeup]{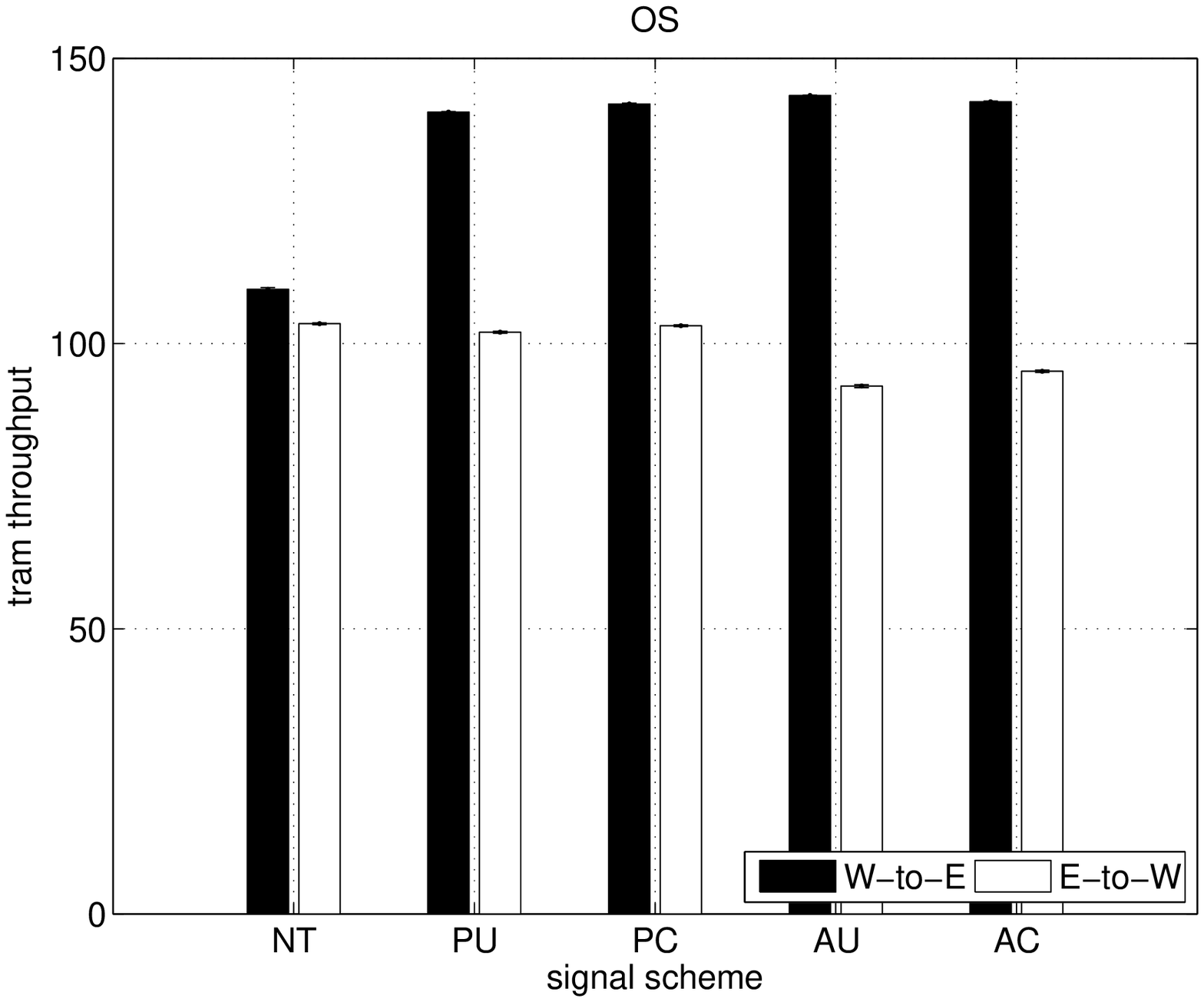}}
\def\tramthroughputUS						{\includegraphics[scale=\threeup]{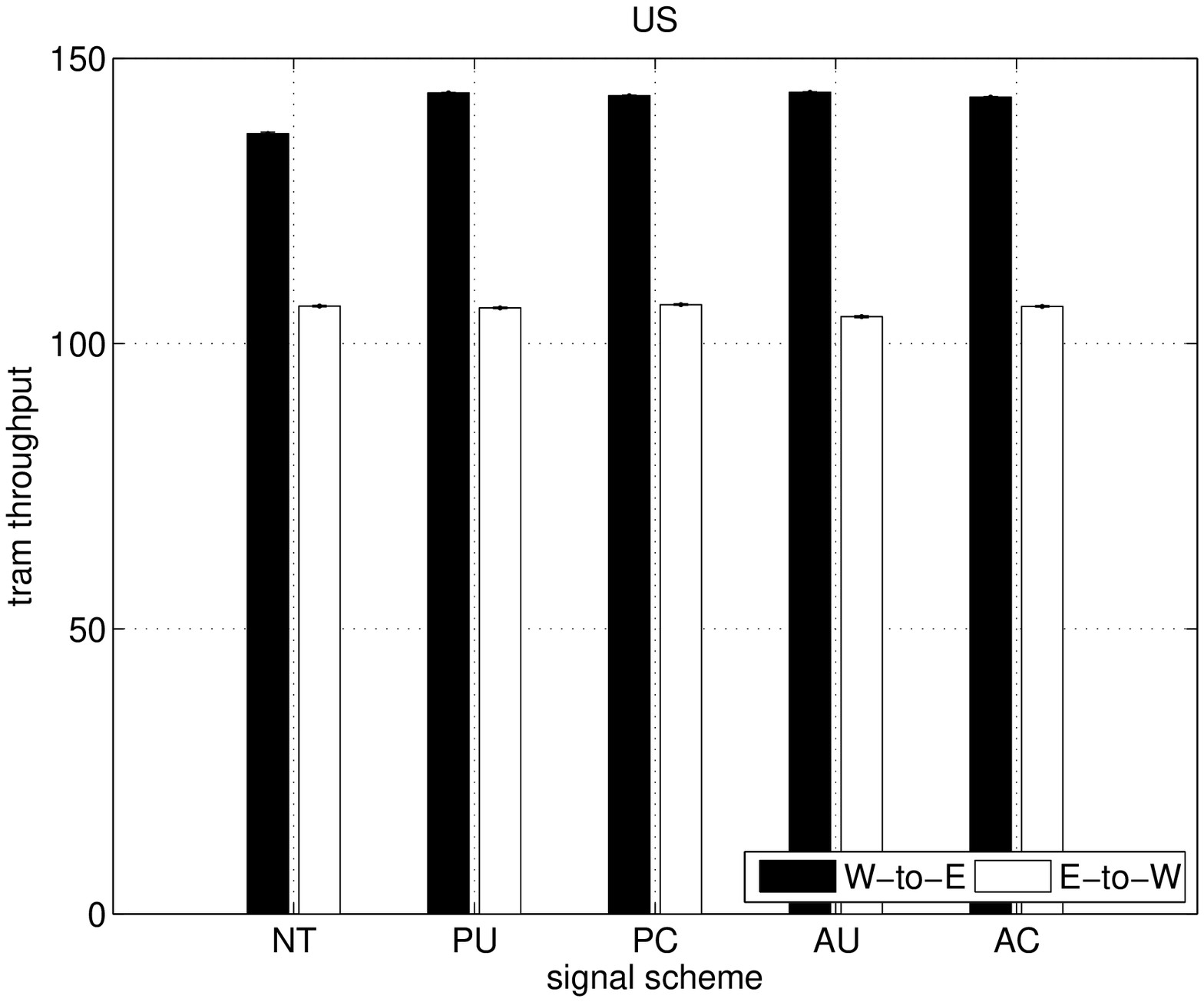}}
\def\tramdevOS								{\includegraphics[scale=\threeup]{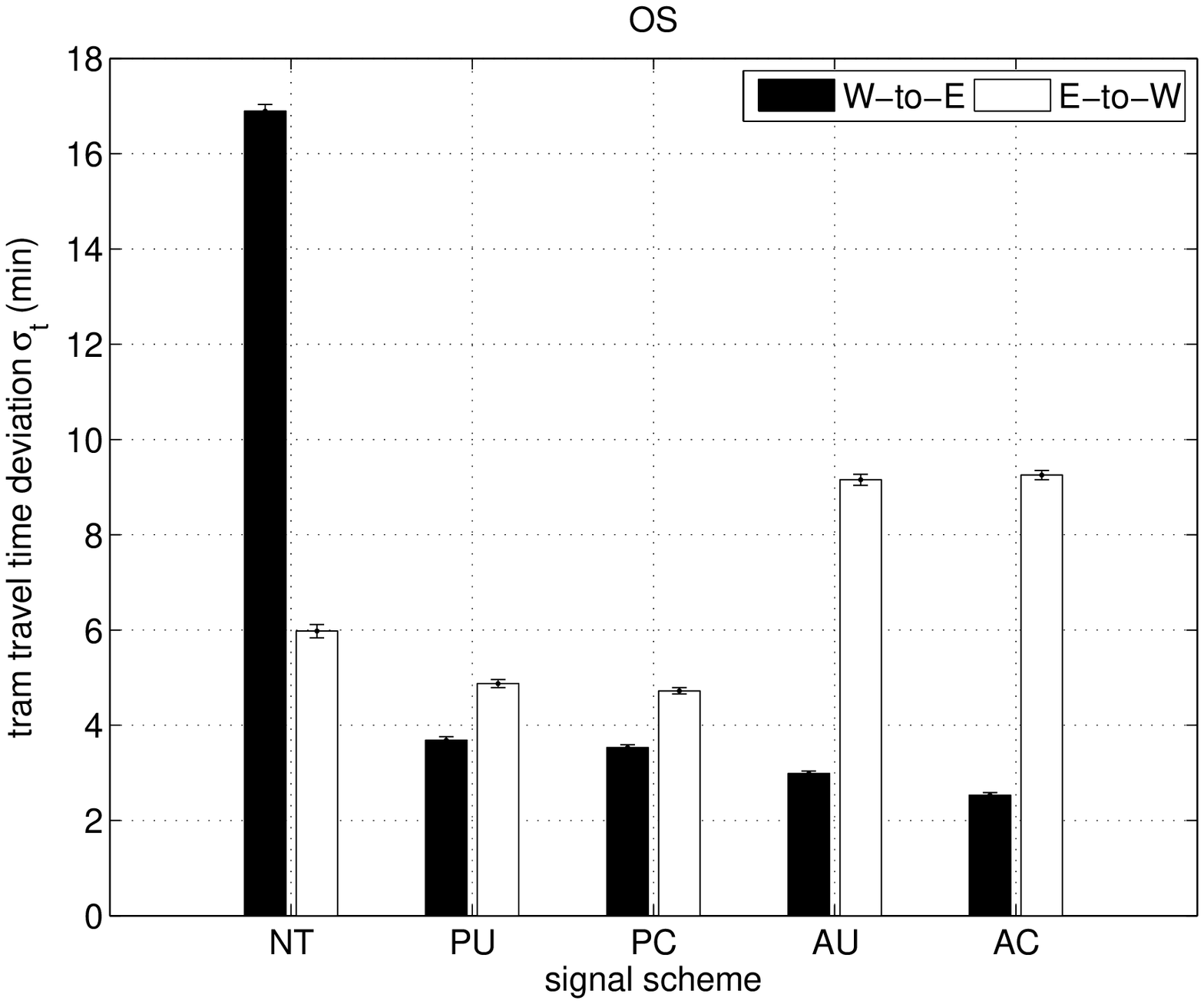}}
\def\tramdevUS								{\includegraphics[scale=\threeup]{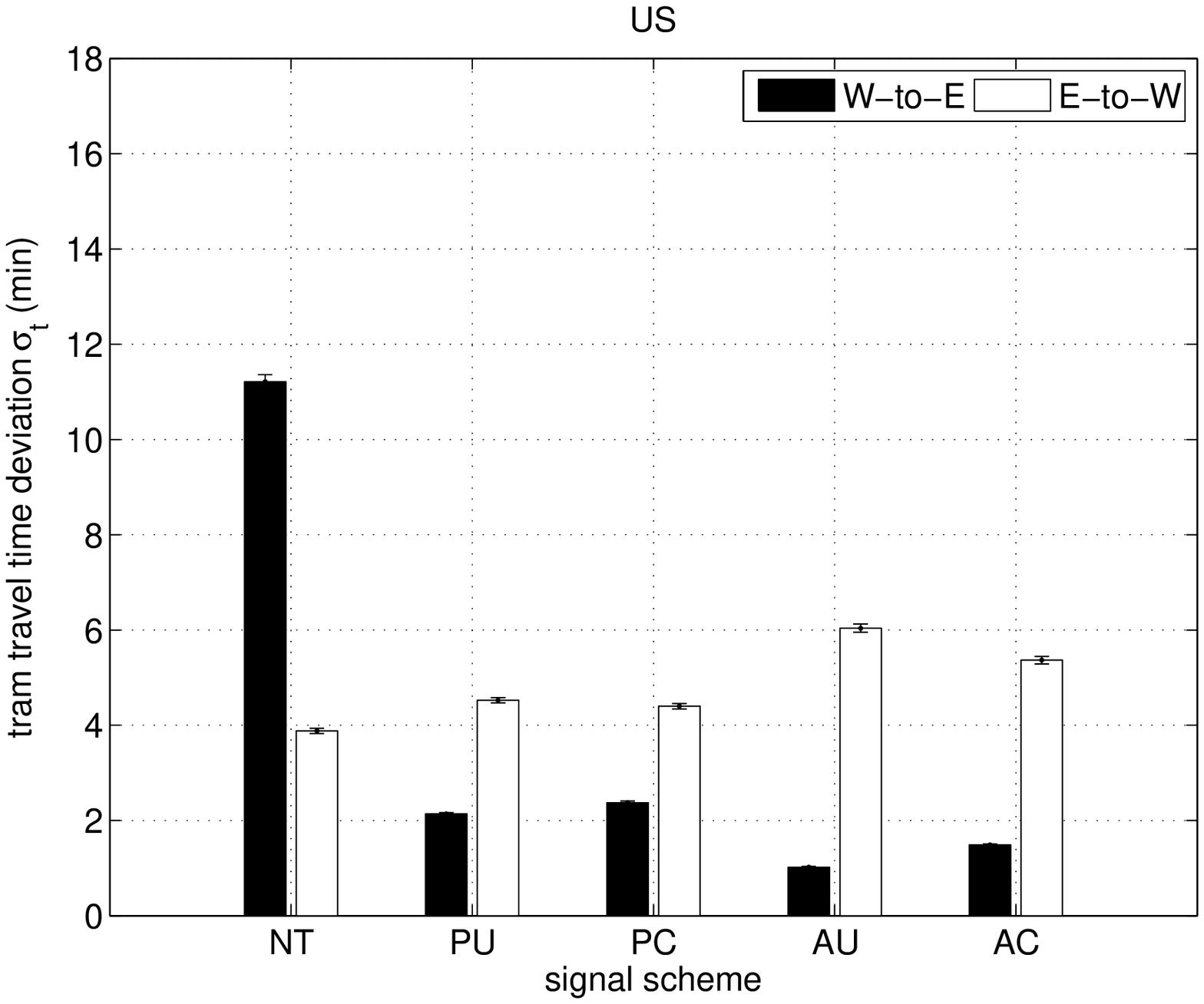}}

\def\carageWEOSN							{\includegraphics[scale=\threeup]{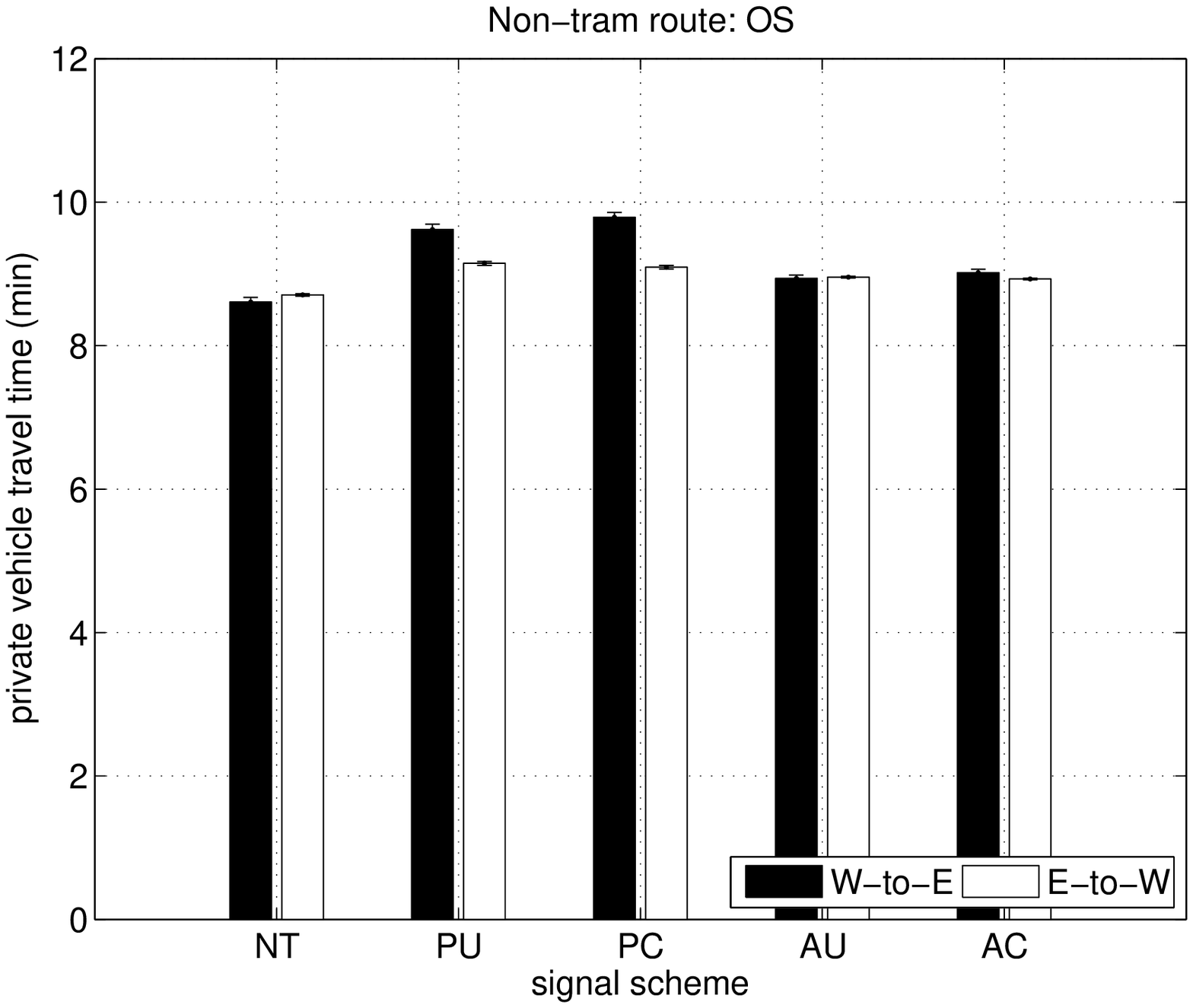}}
\def\carageWEUSN							{\includegraphics[scale=\threeup]{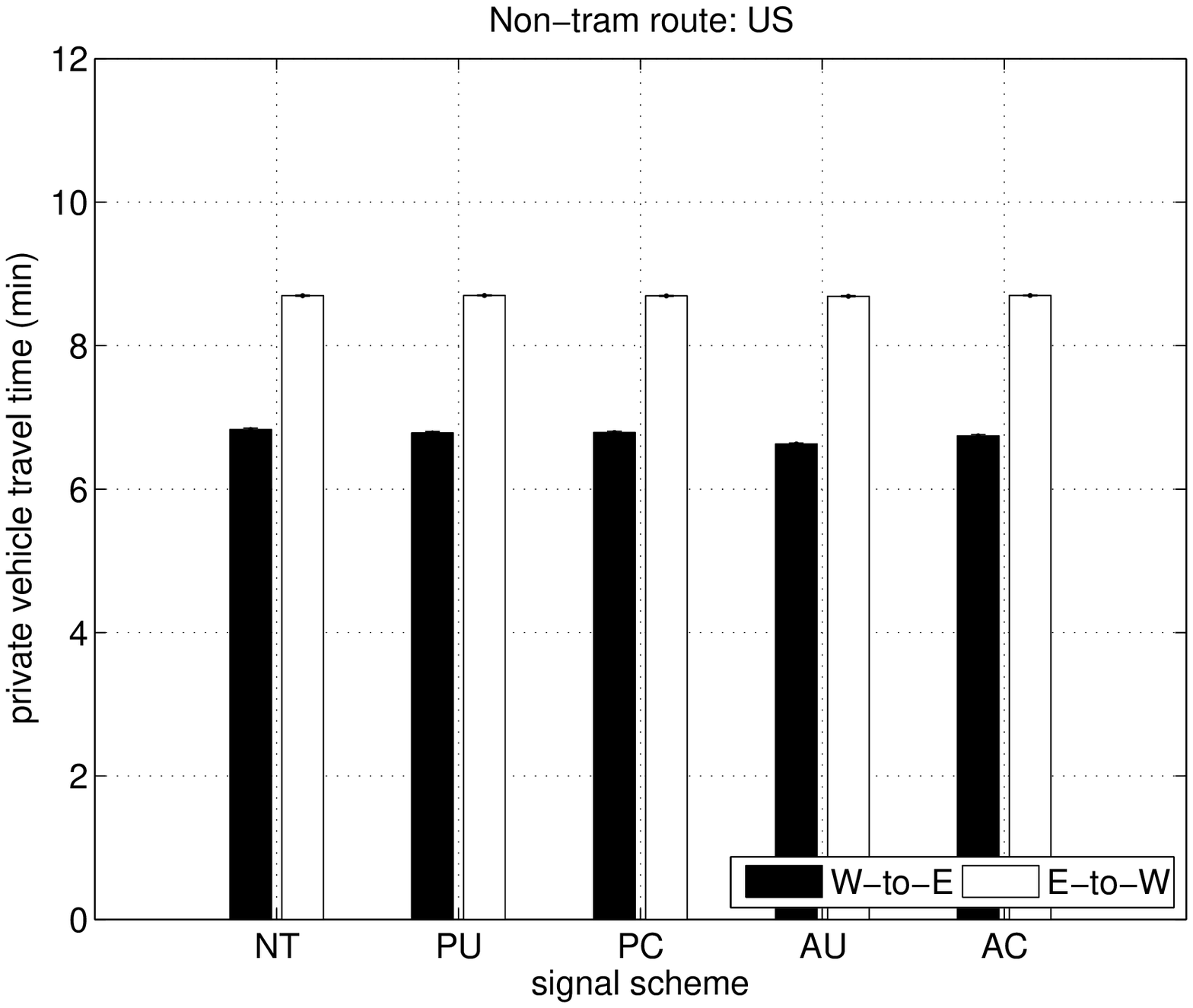}}
\def\carageWEOST							{\includegraphics[scale=\threeup]{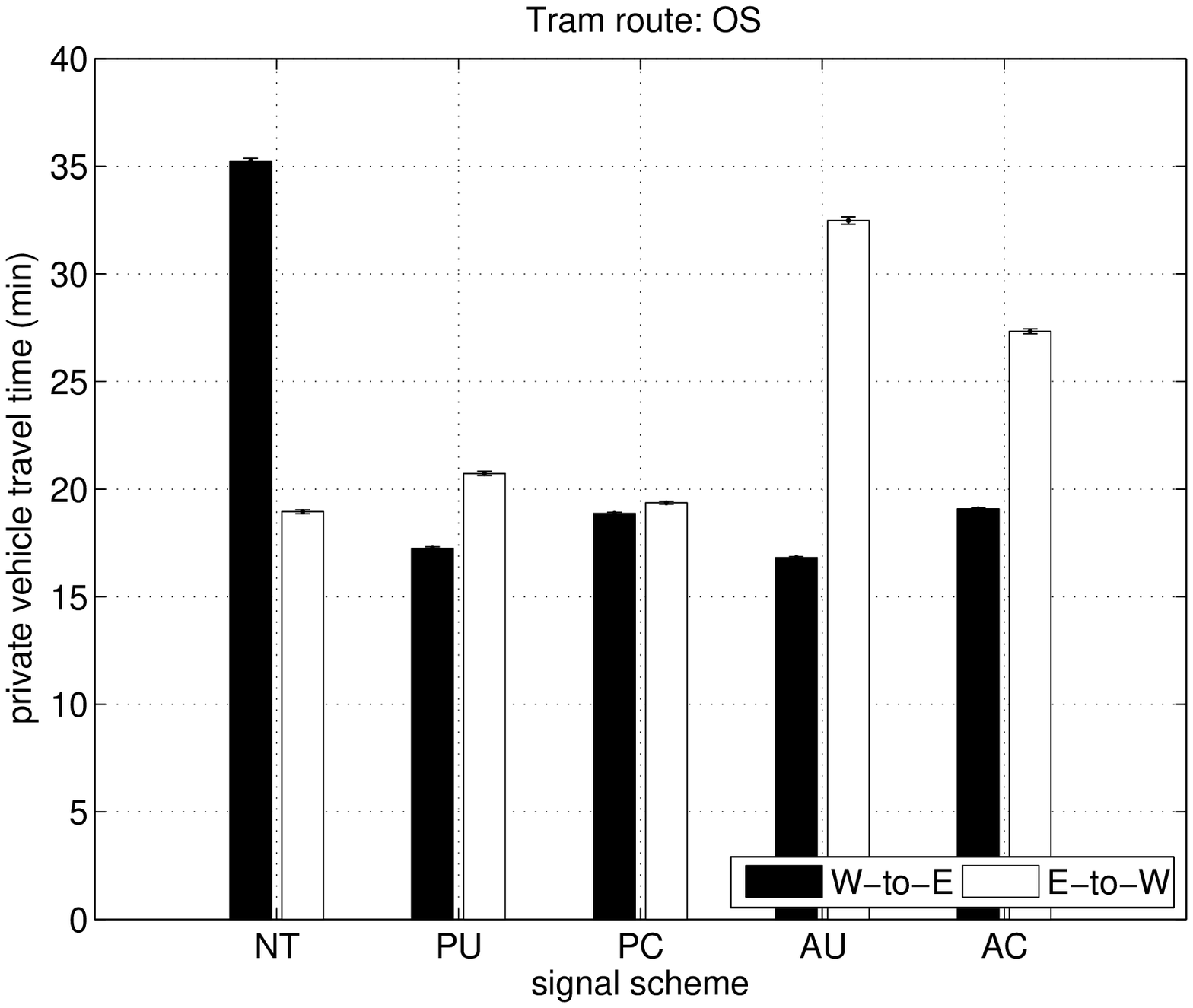}}
\def\carageWEUST							{\includegraphics[scale=\threeup]{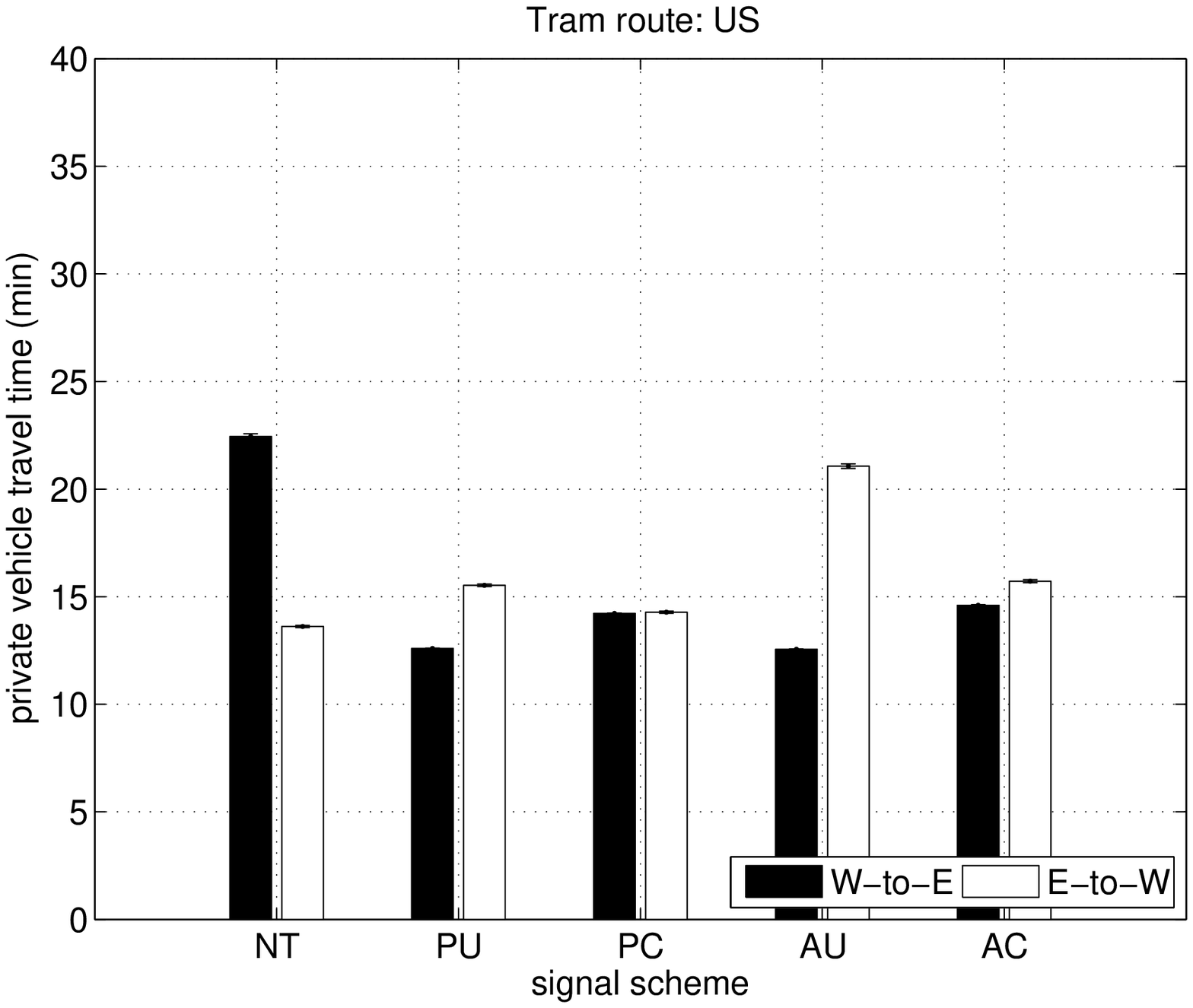}}
\def\carageNSOS								{\includegraphics[scale=\threeup]{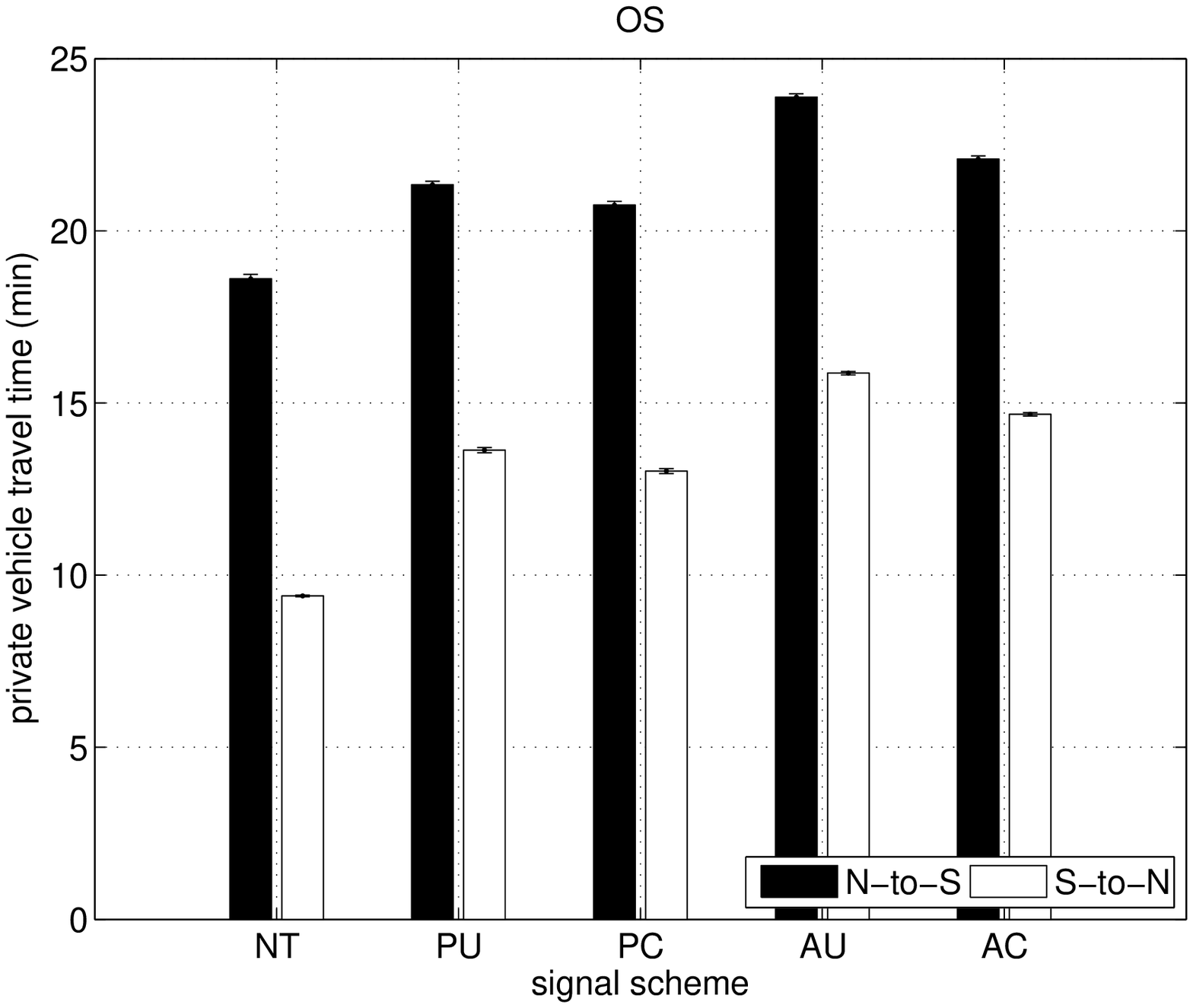}}
\def\carageNSUS								{\includegraphics[scale=\threeup]{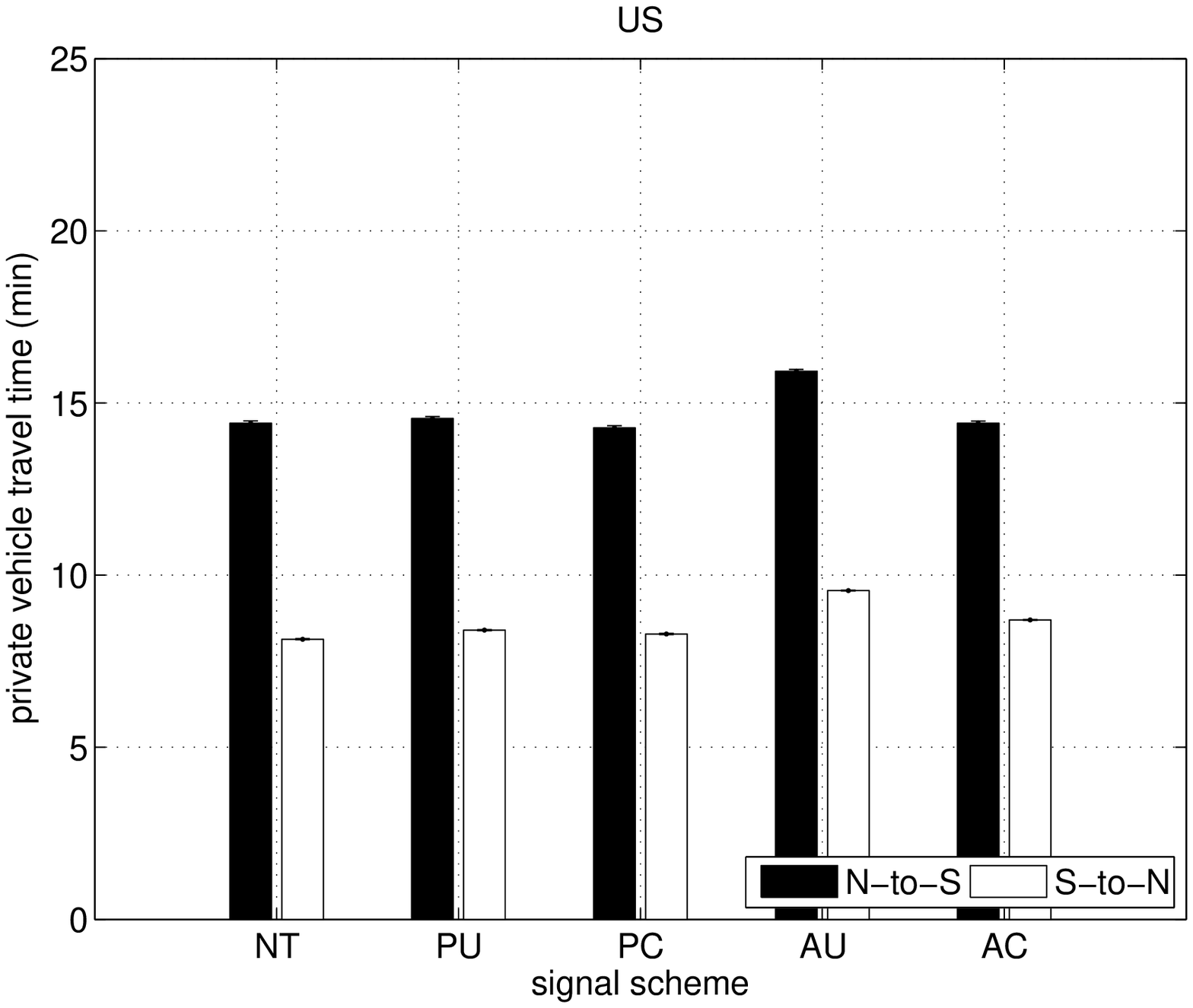}}

\def\peoplethroughputUSALL							{\includegraphics[scale=\threeup]{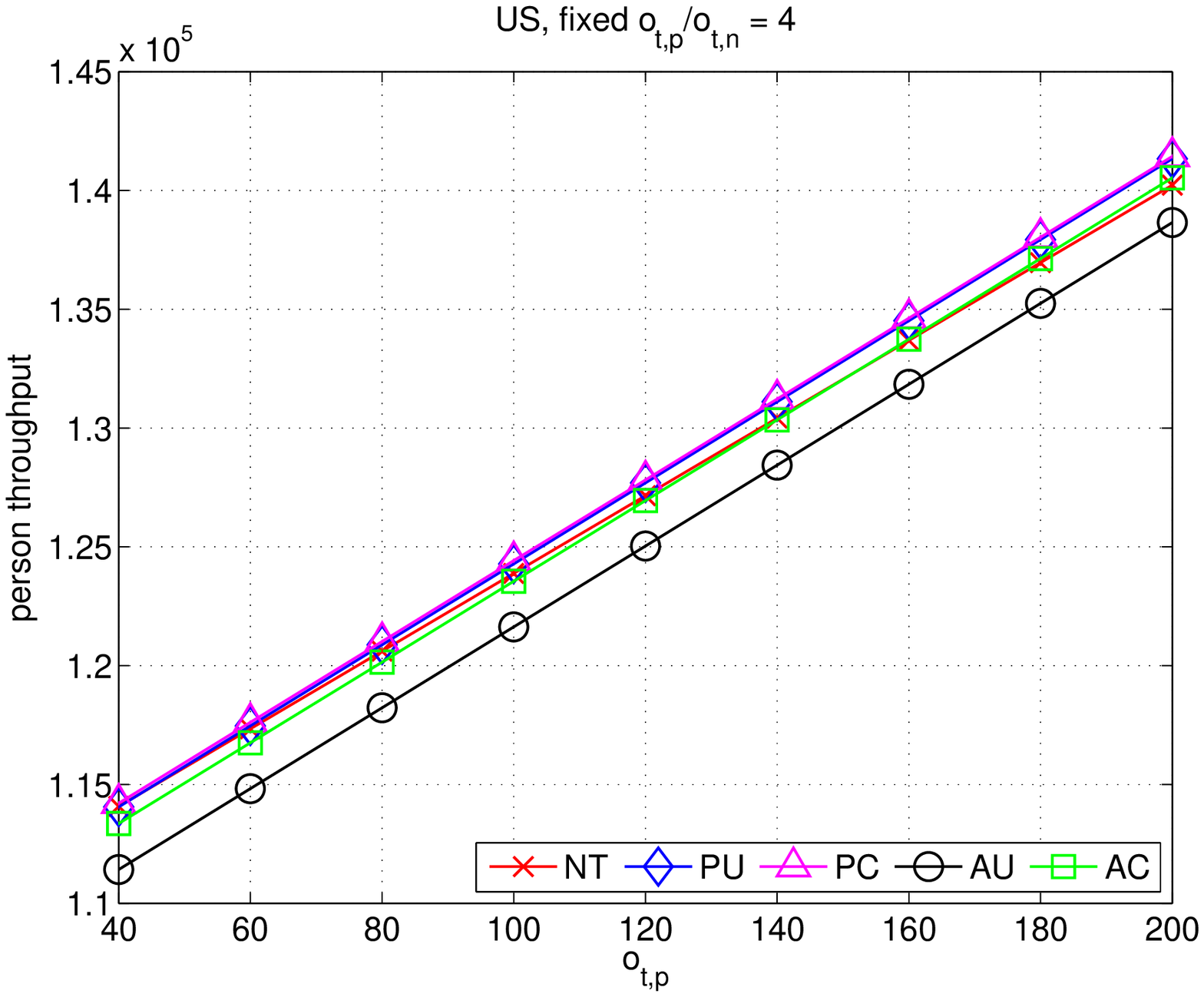}}
\def\peoplethroughputOSALL							{\includegraphics[scale=\threeup]{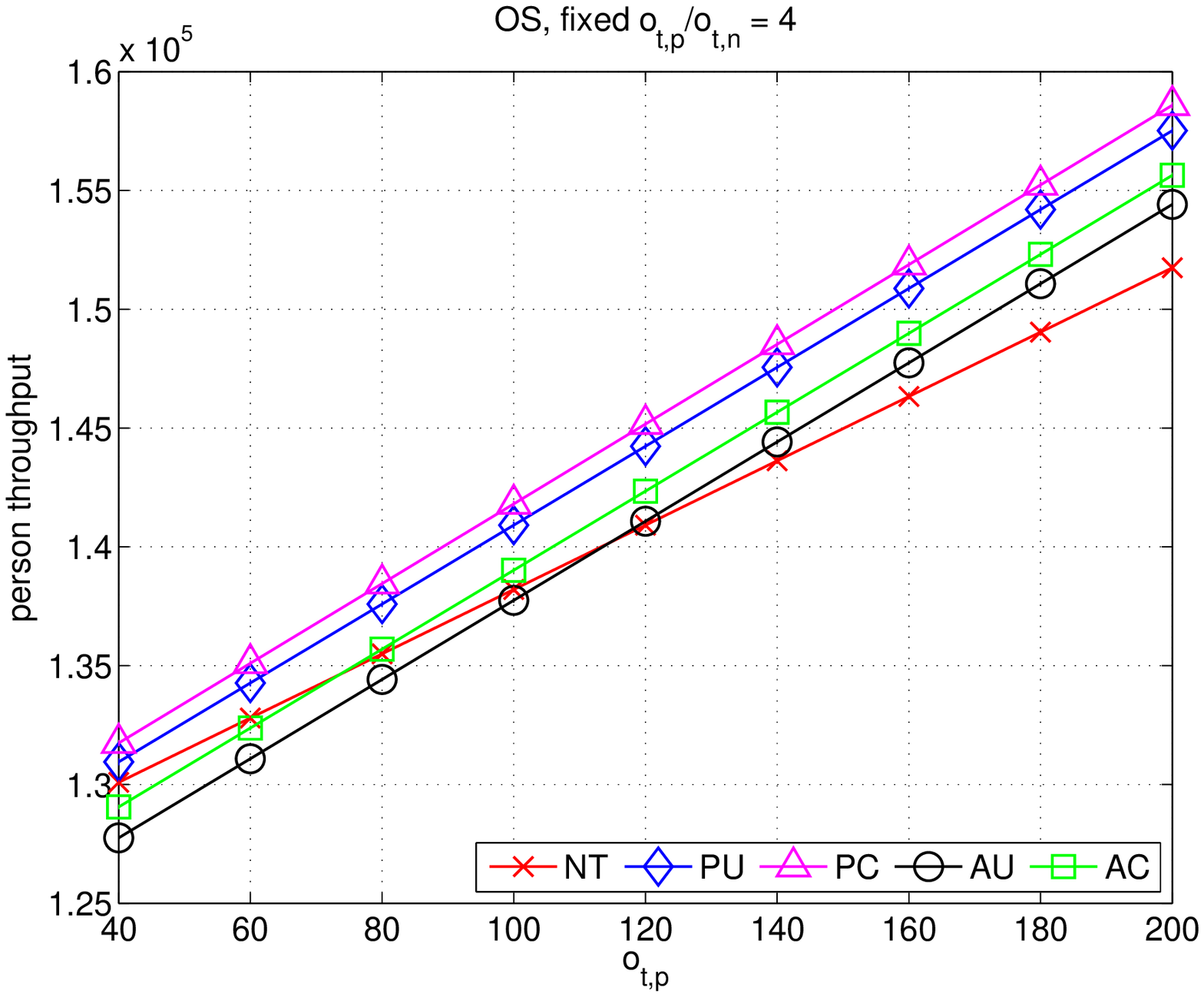}}
\def\peopleageUSALL							{\includegraphics[scale=\threeup]{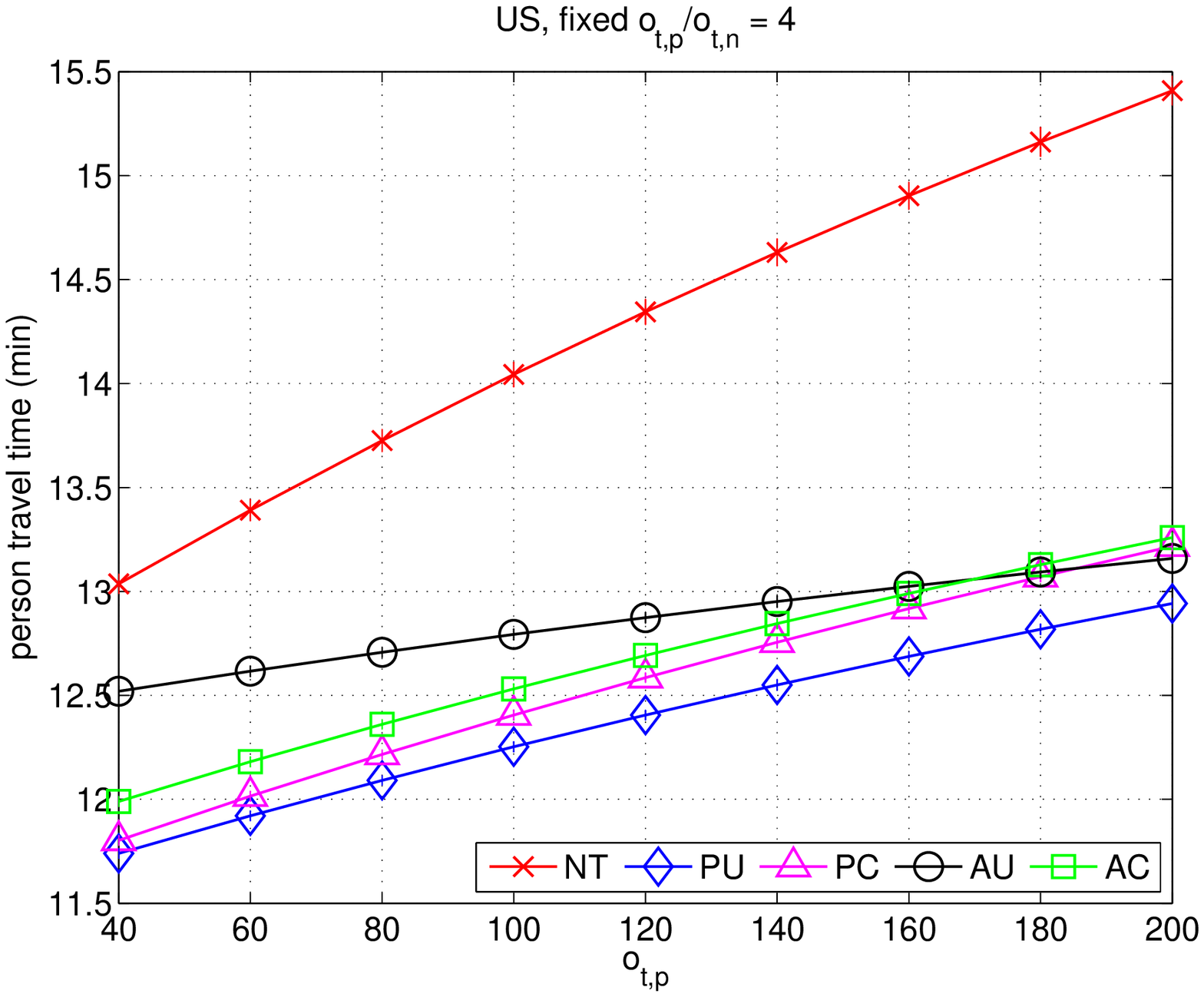}}
\def\peopleageOSALL							{\includegraphics[scale=\threeup]{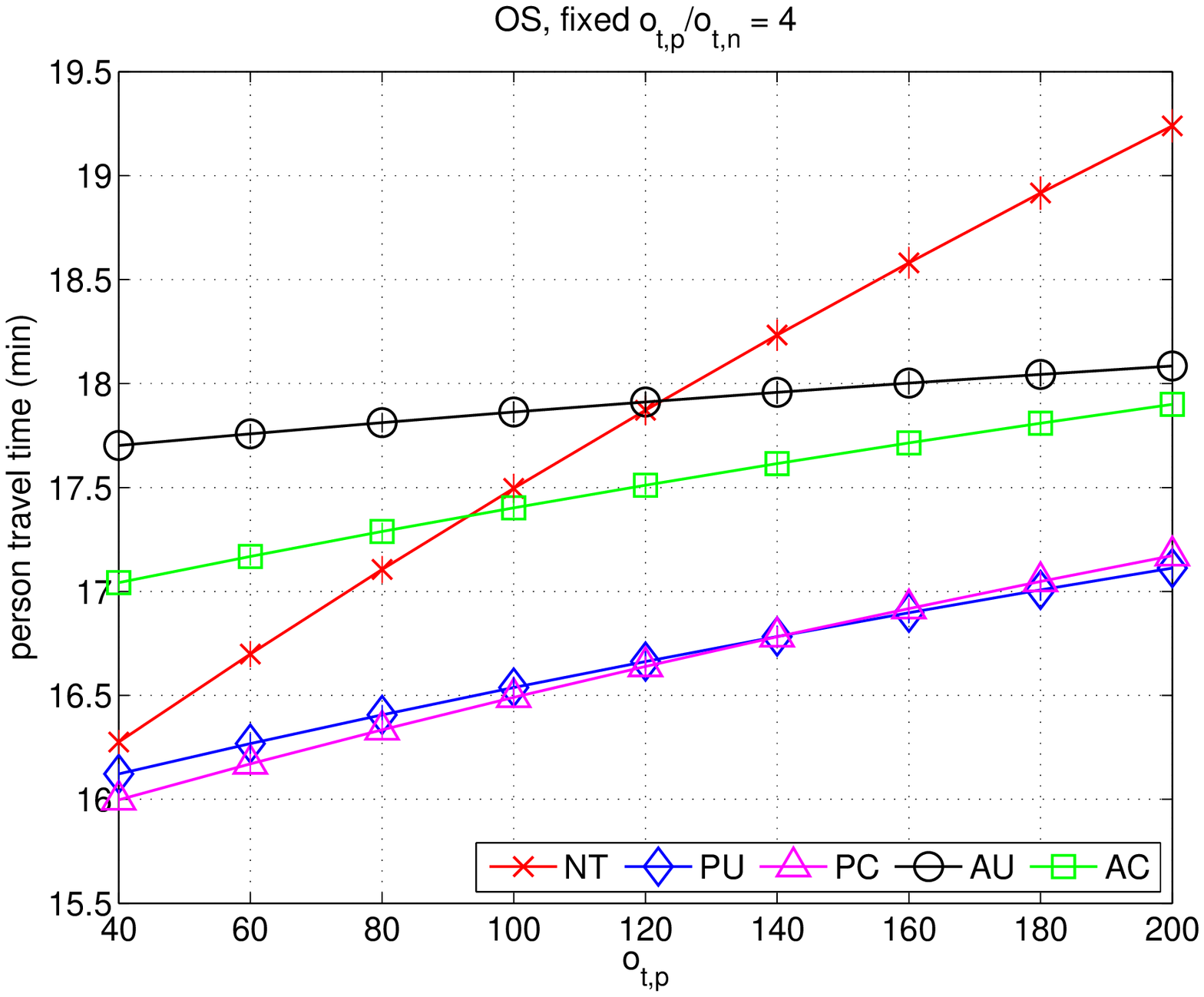}}
\def\peopledevUSALL							{\includegraphics[scale=\threeup]{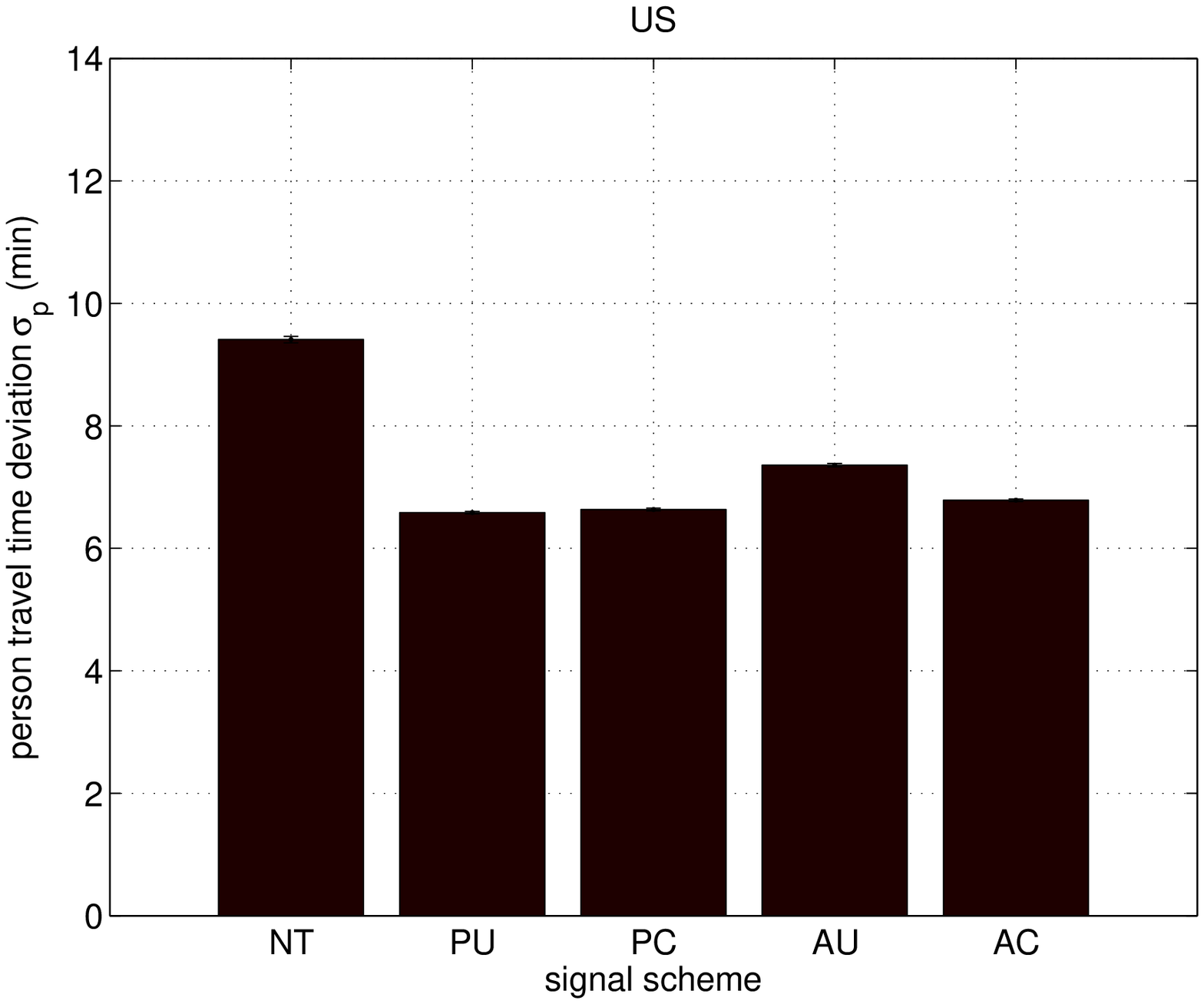}}
\def\peopledevOSALL							{\includegraphics[scale=\threeup]{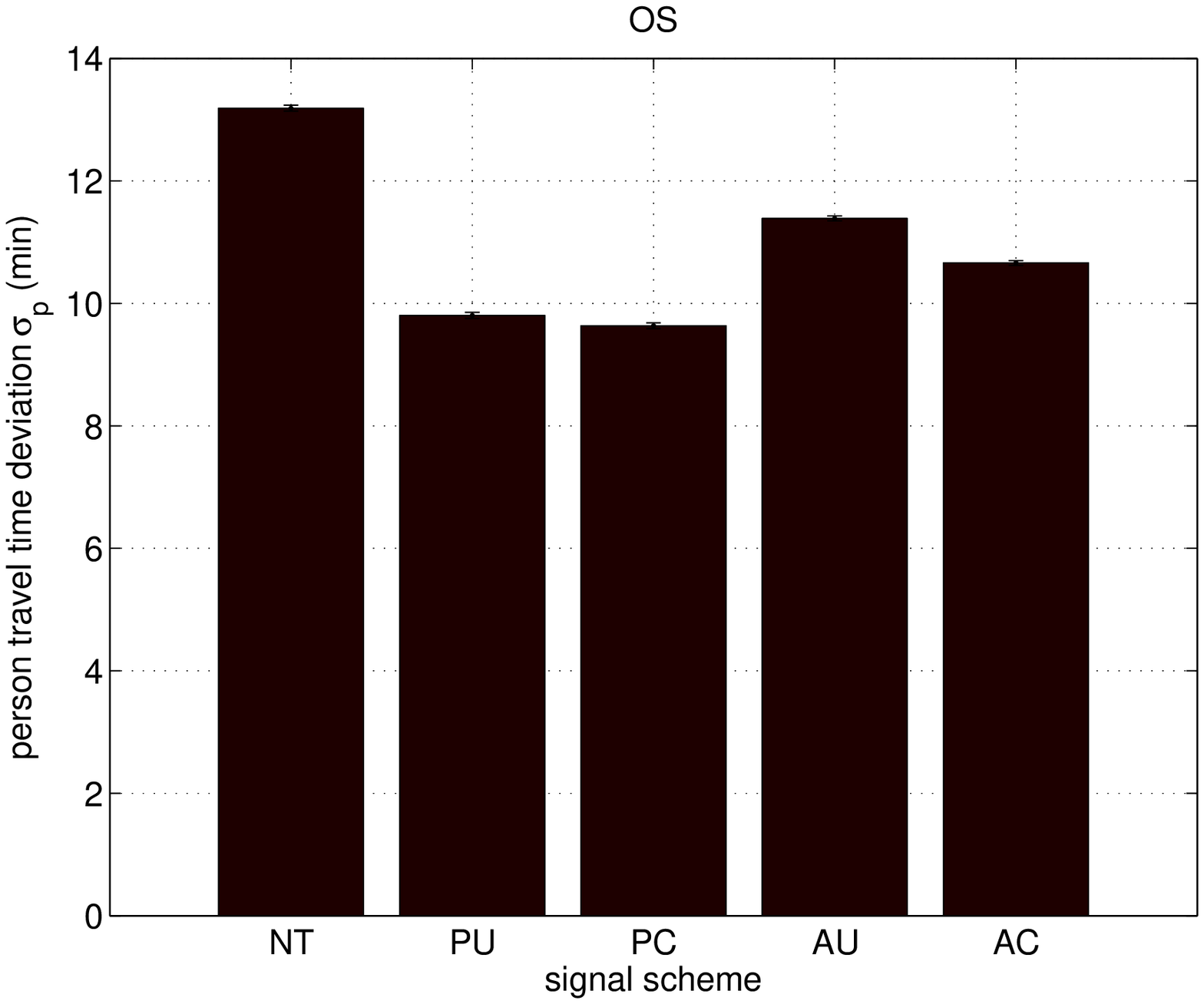}}
\begin{document}

\title{A Comparison of Tram Priority at Signalized Intersections}

\author{Lele Zhang and Timothy Garoni\\
School of Mathematical Science, \\Monash University, Victoria 3800, Australia\\ Tel: +61 3 9905 4503, Fax: +61 3 9905 4403, \\
Email: lele.zhang@monash.edu\\
Email: tim.garoni@monash.edu
}
\maketitle
\begin{abstract}
\normalsize We study tram priority at signalized intersections using a stochastic cellular automaton model for multimodal traffic flow. We simulate realistic traffic signal systems, which include signal linking and adaptive cycle lengths and split plans, with different levels of tram priority. We find that tram priority can improve service performance in terms of both average travel time and travel time variability. We consider two main types of tram priority, which we refer to as full and partial priority. Full tram priority is able to guarantee service quality even when traffic is saturated, however, it results in significant costs to other road users. Partial tram priority significantly reduces tram delays while having limited impact on other traffic, and therefore achieves a better result in terms of the overall network performance. We also study variations in which the tram priority is only enforced when trams are running behind schedule, and we find that those variations retain almost all of the benefit for tram operations but with reduced negative impact on the network.
\end{abstract}


\section{Introduction}

To promote use of public transport, which is a key means of alleviating congestion in urban transport networks, it is important for public transport to run reliably. One useful tool is to provide transit priority at signalized intersections. Transit signal priority (TSP) has been used in practice since the 1970s. Several studies on TSP have been undertaken previously, either via analyzing empirical data \cite{CurrieGohSarvi13,FurthMuller00,KimpelStrathmanBertiniBender04,vanOortvaNes09} or using simulation methods \cite{CurrieSarvi07,MesbahSarviCurrieSaffarzadeh2010,LeeShalabyGreenoughBowieHung2005,HeHeadDing11,JepsonFerreira00}. 

Most of these studies focus on bus signal priority, and very few concern trams. Compared to buses, trams operating in mixed traffic have much higher impact on other road users, {and vice versa.
\begin{itemize}
\item {\it Trams block the entire link when they stop.} For kerbside stops, which are the most typical and common tram stops in Melbourne suburbs \cite{GrahamDennis08,GrahamTivendaleScott11}, when trams stop for loading passengers, they block traffic not only in their own lane but also in adjacent lanes. That is to say, the capacity of the link (at the stop) drops to zero during the loading period. This has a more significant impact when the kerbside stop is at the approach-side of an intersection. Trams stopping at the intersection during a green light will result in a waste of green time. 
\item {\it Trams cannot change lane.} Trams are much more vulnerable to disruptions caused by private vehicles, especially turning vehicles, because they cannot change lane. A single turning vehicle can block a straight-going tram behind it for an entire signal cycle, due to the vehicle's need to give way to oncoming traffic. Such events can cause significant tram delays, even in moderate traffic conditions, unless appropriate tram priority signals are imposed. As a result, tram priority systems usually include a clearance phase to clear turning vehicles in the tram's path. We remark that these events are different from the scenario of buses operating on a separate lane. In Australia, vehicles drive on the left side of the road. Bus lanes are on the left while tram lanes are on the right, which means that even if we assume buses don't change lanes they are unlike to get caught behind turning vehicles because left turners don't give way. 
\item {\it It is comparatively difficult for early trams to stay on schedule.} If buses run ahead of schedule, they usually wait at transit stops, where they will not hinder other road users. However, it is impossible for trams operating in mixed traffic to do this. Trams stopping at kerbside stops will block at least one lane and so reduce the link capacity. Slowing down early trams is also impractical as it leads to capacity reduction. In this paper, besides implementing tram priority signals, we do not consider other strategies for keeping trams on schedule.
\end{itemize}
Due to these specific peculiarities of tram behavior, studies on bus priority cannot be directly adapted to the case of trams.

This paper aims to study the effect of different levels of tram priority via conducting simulations on {\it large-scale networks} governed by {\it realistic adaptive signal systems} with open boundary conditions.} 
Simulation studies of TSP have typically been limited to small-scale networks. \citeasnoun{JepsonFerreira00} studied the impact of active bus priority under various traffic conditions on a 4-lane route. \citeasnoun{CurrieSarvi07} also considered a 4-lane mixed traffic environment. \citeasnoun{HeHeadDing11} evaluated the heuristic algorithm, which deals with multiple requests of priority, on a 2-intersection arterial. \citeasnoun{LeeShalabyGreenoughBowieHung2005} tested the advanced TSP control method on one intersection. These studies were all confined to the question of bus priority. {Furthermore, all of them, except \citeasnoun{CurrieSarvi07}, in which the signal system used was not specified, are confined to the study of fixed cycle signal systems, which are rarely used in practice nowadays.
}

In this paper we utilize a stochastic cellular automaton (CA) model for multimodal traffic networks, to study a number of possible tram priority schemes currently used, or being considered, in Melbourne, Australia. This model was designed with the flexibility to allow the study of multiple vehicle types traveling in an arbitrary multimodal transport network governed by arbitrary signal systems. {As we have mentioned, previous studies using microscopic simulation models focused on small-scale networks, because microscopic simulations require a large amount of input data and time consuming. This CA model is mesoscopic, in the sense that although individual vehicles are modeled, fine-grained details of individual driver behavior are treated in a course-gain, statistical, manner. The model was specifically designed to provide a simple and fast way to study arbitrary traffic signal systems, on large arbitrary networks.} Using this model, we study the behavior of four distinct types of tram priority schemes on a generic 8 by 8 square-lattice network governed by \scats\ (Sydney Coordinated Adaptive Traffic System). 
Of these four types of tram priority, one corresponds to the method currently used by the Melbourne's road authority, VicRoads, while the other three are variations currently being considered for trials. We study and compare the mean tram travel time and variability produced by the different priority systems, as well as the impact on private vehicles. We also evaluate the network performance in terms of person delays. 

Adopting the classification of transit priority strategies mentioned in \cite{FurthMuller00}, the four schemes considered all belong to \em{active} priority, which is to say the signal control system starts priority strategies  when the trams are detected at prescribed locations. The scenarios can be divided into two groups: \em{full} (or \em{absolute}) and \em{partial} priority. Signals with absolute tram priority start the priority phase immediately after detecting a tram and keep the phase running until the tram traverses the approaching intersection. The partial priority group has less disruptive priority tactics, which include a clearance phase and a green extension. For both absolute and partial priority, we consider two variants: \em{unconditional} and \em{conditional} priority. Conditional priority is active only when trams are behind schedule. Unconditional partial priority is the system currently employed in Melbourne. Given the obvious practical importance of improving the reliability and efficiency of public transport, VicRoads is interested in learning under what conditions, if any, absolute tram priority might be desirable. This was the initial motivation for the present study.

\section{Multimodal Traffic Model}

The multimodal CA model used in our simulations extends the NetNaSch unimodal traffic model, (see \cite{deGierGaroniRojas11} for a comprehensive description), to include multimodal traffic and complex vehicular behaviors. The model is designed to simulate large-scale traffic networks with any number of distinct vehicle classes. In this paper, we focus on two vehicle classes of \em{private vehicles} (or \em{cars}) and \em{trams}. 

We now summarize the details of the specific network and input parameters simulated in the present study.

\subsection{Links and Lanes}\label{ssec:links}
\begin{figure}[!t]
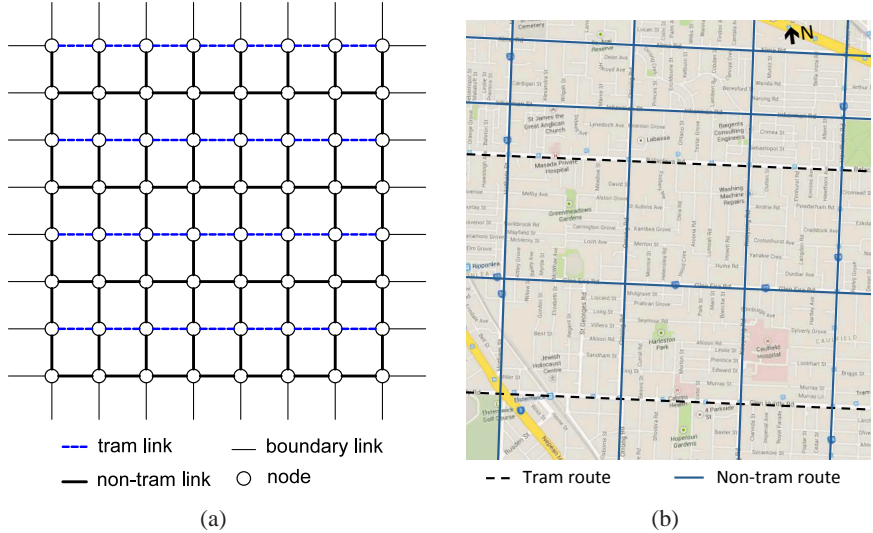

\centering
\subfigure[]{\network} \quad\subfigure[]{\googlemap}
\caption{Left: Illustration of an 8 by 8 square-lattice network studied in our simulations with each alternating east-west route being a tram route. All links carry bidirectional traffic. Boundary links are treated as ramps (buffering zones) for inputting and outputting vehicles. {Right: Google map of the network (main roads only) in St Kilda East and Caulfield, Melbourne, Australia.} }\label{fig:network}
\end{figure}

Melbourne's tram network consists of approximately 250km of track. Of this 250km, approximately 167km of the tracks occur on mixed roadways, in which trams and private vehicles share the same lane. 
The particular network we simulated in this study consists of a regular $8 \times 8$ square grid, illustrated in Fig.~\ref{fig:network}(a). {This is a generic network, however, is also a good representative of typical Melbourne suburban road networks, e.g. see in Fig.~\ref{fig:network}(b).} In this network, each alternating east-west route is a tram route. 
For each tram link there are two lanes, of which the right lane is a car-tram mixed lane, whereas for a non-tram link there are two lanes plus an additional right-turning lane.

{Each link of the model is a simple one-dimensional stochastic CA obeying (a slight generalization of) the Nagel-Schreckenberg dynamics \cite{NagelSchreckenberg92} with simple lane-changing rules. Each lane is discretized into a number of cells, each of 7.5m long, corresponding to the typical space occupied by each private vehicle in a jam. Each vehicle can occupy $z$ cells, $z=1,2,\dots$, and take speed $v=0,1,\dots,v_{\max}$, depending on local traffic conditions. Trams operating in Melbourne vary from 14m to 30m long, and they travel more slowly than private vehicles in mixed traffic. Therefore, in our simulations, we set $z=1$ and $v_{\max} = 3$ for private vehicles, and set $z=3$, i.e., 22.5m long, and $v_{\max}= 2$ for trams. In addition, the model includes, at each time step and for each vehicle, a random unit deceleration which is applied with $p_{\rm noise}$. By appropriately setting $p_{\rm noise}$, we obtain an average free-flow speed for cars of approximately 60km/hr and for trams of 45km/hr (see \cite{deGierGaroniRojas11} for details).}

The length of each bulk and boundary link was set to $750$m, which corresponds to the typical distance between signalized intersections in a suburban road network in Melbourne. The length of each right-turning lane was set to $90$m. The model includes \em{boundary links} as a means of inputting and outputting vehicles, but does not consider them part of the network for the purposes of measuring observables.


\begin{figure}[!t]
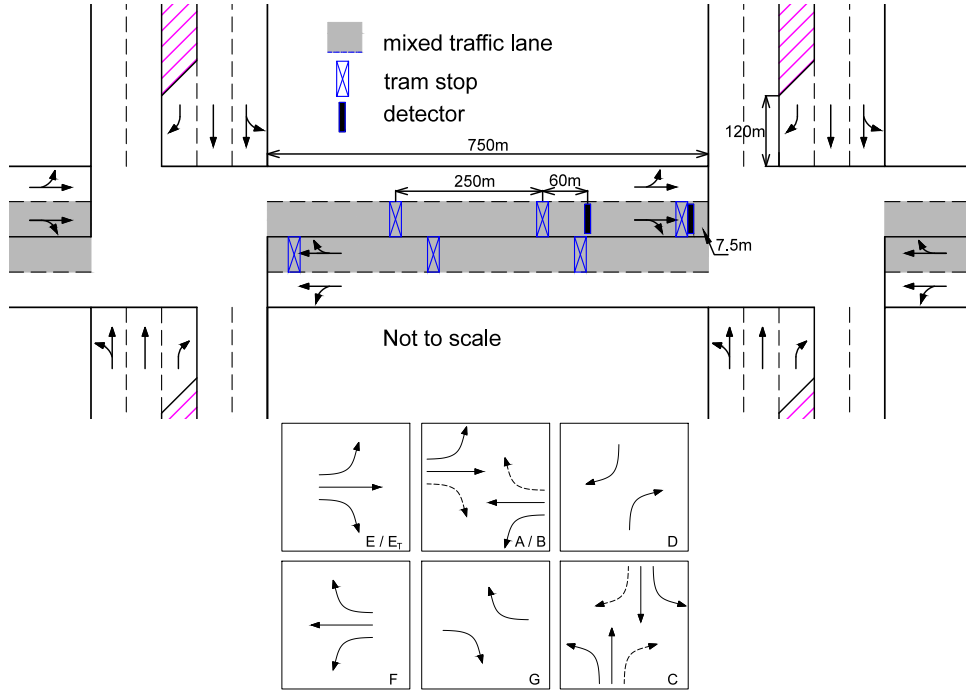

\centering
{\tramIntersection}\\
{\phases}
\caption{{Top: Illustration of tram link. Bottom: }Signal phases. Phases $\ra$, $\rc$, $\rd$, $\rg$, $\rf$ and $\re$ are standard phases, and right-turning clearance phase $\ret$ and extension phase $\rb$ are tram priority phases. Phases $\ra$, $\rc$, $\rd$ and $\rg$ are used at non-tram nodes, whereas phases $\ra$, $\rc$, $\rd$, $\re$ and $\rf$ are used at tram nodes. Dashed paths are required to give way to continuous paths.}\label{fig:phases}
\end{figure}

\rms{Tram Operations}

In actual road networks in Melbourne, a typical scenario along a tram route is that trams operate in the right-hand lane of a two-lane link, with stops being located kerbside. When a tram stops to load/unload passengers, traffic in the left-hand lane must come to a stop in order to give way to passengers. 
In our model, for each tram link there are three stops, one stop every 250m, including one (approach-side) intersection stop and two mid-link stops, as illustrated in Fig.~\ref{fig:phases}. 
In our simulations, the probability $\zeta_s$ that a tram loads/unloads passengers at stop $s$ was set to $1$ if $s$ was an intersection stop and $0.5$ otherwise. The loading/unloading time $\omega_s$ at the intersection stop was $30$ sec, and $20$ sec at the mid-link stops. 

For each tram link in the eastbound (priority) direction, there are 2 tram detectors. The mid-link detector is located 60m after the second mid-link tram stop. The end-link detector is located 7.5m back from the approaching intersection. When a tram passes a detector, the system will register this event and then make an appropriate signal control decision. These control decisions are discussed at length when we describe traffic signal systems in the next section. 

\subsection{Boundary Conditions}\label{ssec:boundary}

In this paper we consider open boundary conditions, and so the density in the network is not controlled directly. Instead, at each time step, vehicles enter a boundary inlink with a prescribed inflow rate and exit via a {boundary outlink} with a prescribed outflow rate. 

We simulated the network over a 4-hour period, and measured the last 3 hours, considering the first hour as a burn-in period. 
We applied two orthogonal peak directions: eastbound and southbound. \scats\ signal coordination (linking) was set in the eastbound direction, to establish green-wave behavior.

\rms{Private Vehicles}

\begin{figure}[!t]
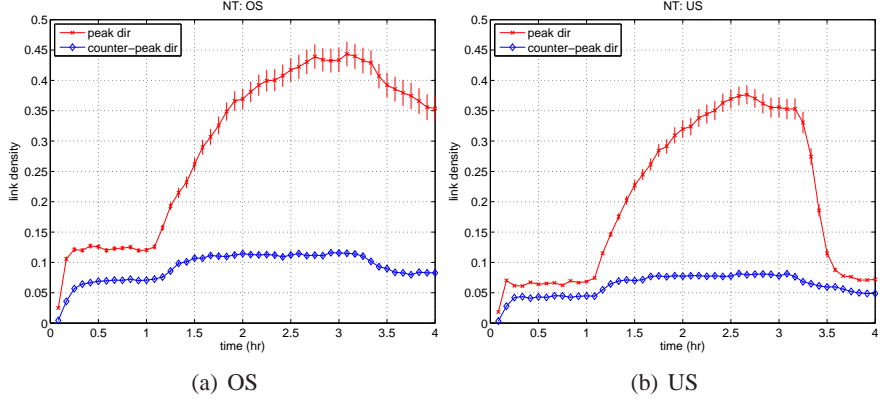

\centering
\subfigure[OS]{\densityprofileOS}\subfigure[US]{\densityprofileUS}
\caption{Time series of densities for two links in the north-to-south (peak) and the south-to-east (counter-peak) directions in the middle of the network with no tram priority.}
\label{fig:inflow}
\end{figure}

The inflow rate for cars follows a typical AM-peak profile, and is higher in the second and the third hours than the other hours. The inflow rates in the peak directions are about twice as large as those in the counter-peak directions during the peak hours. For tram inlinks, the inflow rates of vehicles are only $50\%$ of those for the non-tram inlinks in the same direction. The outflow rates have similar profiles to the inflow rates. We consider two scenarios: over-saturated (OS) and unsaturated (US). {In the US scenario the network is running close to capacity.} Fig.~\ref{fig:inflow} shows the resulting density profiles of two typical links in the north-south direction in the middle of the network when no tram priority system is applied. We note that the precise density profile for each link depends on the signal systems used, the choice of boundary conditions, and the link's location in the network. 
%

%

\rms{Tram Schedules}

Trams are inserted into the network on the boundary inlinks periodically at deterministic times. Every hour 12 trams are scheduled on each tram route in the peak direction, and 9 trams in the counter-peak direction. The first tram starts at time 00:02:00. 

\subsection{Turning Decisions}\label{ssec:turning}

To mimic origin-destination behavior without introducing the computational overhead caused by origin-destination matrices and route planning, the NetNaSch model demands that each cars makes a random choice about its turning decision at the approaching intersection when it enters a link. For trams, those decisions are deterministic as they need to follow routes.

As we consider two peak directions, to match up with the biased boundary conditions, we assumed that turning probabilities for cars are also biased. In our simulations, each link was assigned with a probability $\pT$ of continuing straight ahead, a probability $\pb(1-\pT)$ of turning into a non-peak-direction link, and a probability $(1-\pb)(1-\pT)$ of turning into a peak-direction link. The parameter $\pb$ therefore controls the level of turning bias. 
We used $\pb=0.4$ in our simulations.
The value of $\pT$ depends on the link and node type. For a tram link with no exclusive right turning lane, it is difficult to make a right-turn. In reality, cars try to avoid turning on such links. Also, cars try to avoid turning into tram routes since they do not want to be slowed down and/or frequently stopped by trams. Therefore, we set $\pT=0.85$ at each non-tram node and $\pT=0.9$ at each tram node. 

\subsection{Phases}

Each node in the network was assigned with a set of phases, shown in Fig.~\ref{fig:phases}, depending on the signal system, which will be discussed in the next section. 

\subsection{Observables}

\subsubsection{Throughputs and Travel Times}

In a given simulation, for each value of $\tau$ we have a list $T_{\tau,c}^{(1)},T_{\tau,c}^{(2)},\dots,T_{\tau,c}^{(m_{\tau,c})}$ where $m_{\tau,c}$
is the number (possibly zero) of cars to leave the network at time $\tau$. $T_{\tau,c}^{(i)}$ is the total amount of time spent in the network by the $i$th such car. Analogously, we have a list $T_{\tau,t}^{(1)},T_{\tau,t}^{(2)},\dots,T_{\tau,t}^{(m_{\tau,t})}$ for trams. In a simulation of duration $N$ seconds with measurements starting at $\tau_o$, the total numbers (\em{throughputs}) of cars and trams that have traversed the network are therefore
\begin{align}
\mo_c &:=\sum_{\tau=\tau_o}^{N} m_{\tau,c} \text{ and } \mo_t:=\sum_{\tau=\tau_o}^N m_{\tau,t}.
\end{align}
We define the \em{aggregated travel time} per car and tram by
\begin{align}
\mt_c: &= \frac{\sum_{\tau=\tau_o}^N\sum_{i=1}^{m_{\tau,c}}T_{\tau,c}^{(i)}}{\mo_c}\text{ and } \mt_t: = \frac{\sum_{\tau=\tau_o}^N\sum_{i=1}^{m_{\tau,t}}T_{\tau,t}^{(i)}}{\mo_t}
\end{align}
We also consider the travel time variability, which for trams is given by
\begin{align}
\sigma_t&:=\sqrt{\frac{1}{\mo_t-1}\left(\sum_{\tau=\tau_o}^N \sum_{i=1}^{m_{\tau,t}}\left(T_{\tau,t}^{(i)}\right)^2+\mo_t{\mt_t}^2\right)}.
\end{align}
The value of $\sigma_t$ measures the extent to which the travel time varies from tram to tram on a particular day. Letting $O_{\tau,c}^{(i)}$ ($O_{\tau,t}^{(i)}$) be the number of occupants car $i$ (tram $i$) carries, we further define the throughput of people, travel time per person and person travel time variability by
\begin{align}
\mo_p &:= \sum_{\tau=\tau_o}^{N} \left(\sum_{i=1}^{m_{\tau,c}}O_{\tau,c}^{(i)} +\sum_{i=1}^{m_{\tau,t}}O_{\tau,t}^{(i)}\right),\\
\mt_p &:= \sum_{\tau=\tau_o}^{N} \left(\sum_{i=1}^{m_{\tau,c}}O_{\tau,c}^{(i)}T_{\tau,c}^{(i)} +\sum_{i=1}^{m_{\tau,t}}O_{\tau,t}^{(i)}T_{\tau,t}^{(i)}\right),\\
\sigma_p&:=\sqrt{\frac{1}{\mo_p-1}
\left[\sum_{\tau=\tau_o}^N \left(\sum_{i=1}^{m_{\tau,c}}O_{\tau,c}^{(i)}\left(T_{\tau,c}^{(i)}\right)^2+ \sum_{i=1}^{m_{\tau,t}}O_{\tau,t}^{(i)}\left(T_{\tau,t}^{(i)}\right)^2\right)+\mo_p{\mt_p}^2\right]}.
\end{align}

In our simulations, we assumed that the number of occupants that each car carries is identical, that is, $O_{\tau,c}^{(i)} = o_c=1.2$ \cite{VicroadsReport2013}. Furthermore, we assumed that the occupancy of trams operating in the same direction is the same. Namely, $O_{\tau,t}^{(i)}=o_{t,p}=80$ if the tram runs in the peak direction and otherwise $O_{\tau,t}^{(i)}=o_{t,n}=20$. We shall discuss the person performance as a function of $o_{t,p}$ and $o_{t,n}$ with a fixed ratio $o_{t,p}/o_{t,n}=4$.

\subsubsection{Statistics}

For each distinct choice of traffic signal systems and boundary conditions, we performed $100$ independent simulations, in order to estimate the expected values of the quantities defined in the last subsection. We used one standard error to set the error bars.

\section{Traffic Signal Systems}\label{sec:signals}

The \scats\ traffic signal system uses knowledge of the recent state of traffic to choose appropriate values of three key signal parameters: cycle length, split time, and linking offset. At each intersection it can adaptively adjust both the total cycle length, and the fraction (\em{split}) of the cycle given to each particular phase. In addition, it can coordinate (\em{link}) the traffic signals of several consecutive nodes along a predetermined route in a subsystem by introducing \em{offsets} between the starting times of specific phases, thereby creating a green wave. Since all of the tram priority process discussed here are/would be implemented in Melbourne using \scats, we now give a brief description on the relevant details of our model of \scats.

{\subsection{Signal Linking}\label{ssec:linking}

A \em{subsystem} is a group of nodes which all share a common cycle length.
Within each subsystem, we appoint a unique \em {master} node $\rm m$, and the remaining nodes are \em{slave} nodes. The common cycle length of the subsystem is determined by the master node based on its local traffic condition.
The plot in Fig.~\ref{fig:linking} illustrates a subsystem on an east-west route, which is used in our simulation. 

\begin{figure}[!t]
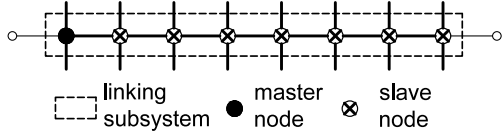

\centering
{\linking}
\caption{Illustration of a linking subsystem.}
\label{fig:linking}
\end{figure}

To implement linking, each node is assigned a special phase $\mp^*$, which is its \em{linked phase}. The linked phase of a slave node $\rm s$ is coordinated to start $\delta$ sec after that of the master node $\rm m$. 
Ideally, the linking offset $\delta$ should be chosen based on the distance $L$ between $\rm m$ and $\rm s$, and the instantaneous local space-mean speed.
In practice, actual implementations of \scats\ tend to operate with fixed offsets during a given period of the day (for example morning peak hour). 
In our simulations we therefore used a fixed \em{linking speed} $52$km/hr, which is just slightly less than the average free-flow speed of 60km/hr.}

\subsection{Adaptive Cycle Length and Split Plan}\label{ssec:sp}

{In practice,} \scats\ chooses cycle length $\mc$ based on local traffic conditions, as quantified by the \em {Degree of Saturation} (\ds). {If traffic is congested, signaled by a large \ds, then cycle length is increased by a fixed amount. Conversely, if green time was wasted during the previous cycles, signaled by a small \ds, cycle length is decreased.} In our model, \ds\ was estimated using stop-line occupancy and flow through the intersection. Cycle length varies from $48$sec to $134$sec. See Appendix for the detailed \scats\ cycle length algorithm used in our model.

Once a new cycle length $\mc'$ is determined, the new split time $\ms_\mp'$ of phase $\mp$ is taken to be proportional to $\ms_\mp\ds_\mp$ of the previous cycle: 
\begin{align}
\ms_\mp'&=\frac{\ms_\mp\ds_\mp}{\sum_{\mp} \ms_\mp\ds_\mp}[\mc'-\mbox{number of phases}\times\ms_{\min}-\mbox{total amber time}] + \ms_{\min}.\label{equ:split}
\end{align}
In the case where some phases are imposed with fixed split, the above expression can be easily modified to choose split times for the remaining phases. An amber time of $2$ sec is imposed for each phase-swap unless the two phases have precisely the same set of paths, e.g. $\ra$ to $\rb$. A minimum split time $\ms_{\min}=5$ sec for non-tram-priority phases was used in our simulations.

\subsection{Versions of \scats}\label{ssec:schemes}

On non-tram nodes, whose inlinks are all non-tram links, we apply the above \scats\ model with signal linking from east to west and phases $\ra$, $\rd$, $\rc$ and $\rg$ in Fig.~\ref{fig:phases}. For tram nodes, we consider five variants of \scats\ with/without tram priority:
\begin{itemize}
\item[\nt.] \scats\ with no tram priority. 
\item[\tu.] \scats\ with partial and unconditional tram priority. 
\item[\tc.] \scats\ with partial and conditional tram priority. 
\item[\au.] \scats\ with absolute and unconditional tram priority. 
\item[\ac.] \scats\ with absolute and conditional tram priority. 
\end{itemize}

We do not apply linking along tram routes since the tram priority phase and tram loading/unloading renders the linking inefficient. 
Therefore tram nodes choose their own cycle lengths and split plans according to their local traffic conditions, independent of their neighbors.

\nt\ is the basic \scats\ system for tram nodes with no tram priority. {In short, it assigns $20\%$ of the cycle length to either phase $\re$ or $\rf$, and runs $\rf$ once and $\re$ twice every $3$ cycles. Cycle length and split plan are chosen prior to the start of each cycle, which is independent of the status of trams, and are not modified mid-cycle. The mechanism of \nt\ is given in Algorithm~\ref{alg:nt}.

\begin{myalg}{\nt}\label{alg:nt}

\smallskip
\noindent
\begin{tabular}{p{11.5cm}}
\hline
{\bf if } node $\rm n$ is to restart a new cycle, {\bf then}\\
\hspace{0.5cm} Increment $c({\rm n})$\\
\hspace{0.5cm} Choose $\mc$ using Algorithm~\ref{alg:scats}\\
\hspace{0.5cm} {\bf if } $c({\rm n})\%3 \neq 0$, {\bf then}\\
\hspace{1cm} Set $\ms_{\re} = 20\%\mc$\\
\hspace{0.5cm} {\bf else}\\
\hspace{1cm} Set $\ms_{\rf} = 20\%\mc$\\
\hspace{0.5cm} {\bf end if}\\
\hspace{0.5cm} Choose splits for $\ra$, $\rc$ and $\rd$ using \eqref{equ:split} with the remaining cycle time\\
{\bf else}\\
\hspace{0.5cm} Implement phases $\re$ (or $\rf$), $\ra$, $\rc$ and $\rd$ in order\\
{\bf end if}\\
\hline
\end{tabular}

*Function $x\%y$ gives the remainder from dividing $x$ between $y$.
\end{myalg}

The observable $c({\rm n})$ acts as a counter for node $\rm n$, recording how many cycles have been implemented.} 
The purpose of phases $\re$ and $\rf$ is primarily to clear right-turning cars in the east-west direction. One of them is implemented each cycle. Since the right lane is shared by trams and cars, it is difficult to obtain a good estimate of \ds\ in this lane. In actual practice, a fixed split is therefore imposed on $\re$ (and $\rf$). 

\smallskip
If no tram priority process is active, then \tu, \tc, \au\ and \ac\ behave essentially the same as \nt. When a tram passes a mid-link detector, a priority process is called provided that no one is already running. Then priority phases are implemented according to the location of the tram. {To simplify the exposition of the priority process, we use variable $\Delta$ to indicate the position of the tram that has triggered the process: $\Delta = 1$ if the tram has passed the middle-link detector but not the end-link one; $\Delta = 2$ if the tram has passed the end-link detector but has not traversed the intersection; and $\Delta = 0$ if the tram has traversed the intersection.}

{\tu\ includes two tram priority phases $\ret$ and $\rb$, and runs $\ret$, $\re$ (or $\rf$), $\ra$, $\rb$, $\rc$ and $\rd$ in order, possibly skipping $\ret$ and/or $\rb$. In short, when $\Delta=1$, \tu\ runs phase $\ret$ to let cars, especially those turning right, traverse the intersection and clear a passage ahead of the tram. When $\Delta=2$, \tu\ runs phase $\rb$ in order to increase the probability that the tram traverses the intersection in the current cycle.} The detailed signaling algorithm for \tu\ when priority process is active is given in Algorithm~\ref{alg:tu}.


\begin{myalg}{\tu}\label{alg:tu}

\smallskip
\noindent
\begin{tabular}{p{11.5cm}}
\hline
{\bf if } node $\rm n$ is to restart a new cycle, {\bf then}\\
\hspace{0.5cm} {\bf if} phase $\rf$ is to run, {\bf then}\\
\hspace{1cm} Replace $\rf$ with phase $\re$\\
\hspace{0.5cm} {\bf end if}\\
\hspace{0.5cm} Subtract as much as $20\%\mc$ from the nominal split time of phase $\rc$
\\
\hspace{0.5cm} {\bf if} $\Delta = 1$, {\bf then}\\
\hspace{1.cm} Give $15\%\mc$ to phase $\ret$ and $5\%\mc$ to phase $\rb$\\ 
\hspace{0.5cm} {\bf else} ({\bf if } $\Delta = 2$, {\bf then})\\
\hspace{1.cm} Give $20\%\mc$ to phase $\rb$\\
\hspace{0.5cm} {\bf end if}\\
{\bf else}\\
\hspace{0.5cm} {\bf if} phase $\ret$ is running and $\Delta = 2$, {\bf then}\\
\hspace{1cm} Terminate phase $\ret$ immediately, skip phase $\re$ and initiate phase $\ra$\\
\hspace{0.5cm} {\bf end if}\\
\hspace{0.5cm} {\bf if} phase $\rb$ is to start and $\Delta = 1$, {\bf then}\\
\hspace{1cm} Skip phase $\rb$\\
\hspace{0.5cm} {\bf end if}\\
\hspace{0.5cm} {\bf if} $\Delta = 0$, {\bf then}\\
\hspace{1cm} Terminate (or skip) phase $\ret$ (or $\rb$) immediately\\
\hspace{0.5cm} {\bf end if}\\
\hspace{0.5cm} {\bf if} phase $\ret$ and/or $\rb$ is terminated early or is skipped, {\bf then}\\
\hspace{1cm} Return the unused time to phase $\rc$\\
\hspace{0.5cm} {\bf end if}\\
{\bf end if}\\
\hline
\end{tabular}
\end{myalg}


\tu\ is a partial priority system in the sense that the time for running priority phases $\ret$ and $\rb$ is limited, not more than $20\%\mc$ per cycle, and the implementation of $\ret$ and $\rb$ may suffer delays. For \tu\ the priority process can be called at any time during a cycle, but will not take effect until the following cycle. This delay may result in efficiency reduction of the priority process. Next we proceed to the signal system with absolute tram priority, which has no delay or restriction on time in implementing the priority phase.

Unlike \tu, the \au\ system does not run the extension phase $\rb$. When the tram priority process is triggered, it starts phase $\ret$ {\it immediately} and keeps running it indefinitely until the tram that triggered the process has traversed the intersection. The detailed algorithm is given in Algorithm~\ref{alg:au}.

\begin{myalg}{\au}\label{alg:au}

\smallskip
\noindent
\begin{tabular}{p{11.5cm}}
\hline
Let $\mp_{i}$ be the phase that is interrupted by the priority process \\
Terminate $\mp_{i}$ immediately provided that it has run for $\ms_{\min}$\\
{\bf if} $\Delta \neq 0$, {\bf then}\\
\hspace{0.5cm} Run phase $\ret$ \\
{\bf else}\\
\hspace{0.5cm}{\bf if} $\mp_{i+1} = \re$, where $\mp_{i+1}$ is the phase following $\mp_i$, {\bf then}\\
\hspace{1.cm} Run phase $\rf$\\
\hspace{0.5cm}{\bf else} \\
\hspace{1.cm} Run $\mp_{i+1}$\\
\hspace{0.5cm} {\bf end if}\\
{\bf end if}\\
\hline
\end{tabular}
\end{myalg}

For \au\ any phase may be terminated early. To avoid possible pathological \ds\ values caused by the priority process, the cycle length and the split plan for next cycle are not updated unless no phase is closed earlier than it should be.

\smallskip
\tc\ and \ac\ are conditional variants of \tu\ and \au\ respectively. In these cases, tram priority processes can be called only if trams are behind schedule. In order to determine whether a tram is running on time, we assign each detector $d$ with an expected arrival time ${\bar T}_d$. A tram is then considered to be late if its travel time when it arrives at the detector $d$ is larger than ${\bar T}_{d}$. The expected arrival time ${\bar T}_d$ is computed based on the location of the detector and the expected travel speed of the tram, excluding loading time. Let $L_d$ be the travel distance to the detector $d$ and ${\bar v}_t$ be the expected tram travel speed. The expected accumulated tram loading time is ${\bar P}_d =\sum_{i=1}^s \zeta_i\omega_i$, where $i=1,2,\dots, s$ are stops that the tram has passed so far and $\zeta_i$ and $\omega_i$ are the stopping probability and loading time at stop $i$. Therefore ${\bar T}_d= \frac{L_d}{{\bar v}_t}+{\bar P}_d$. We used ${\bar v}_t = 27$km/hr in our simulations. {The expected travel speed including loading time is then about $18$km/hr, which is consistent with the average tram travel time in the mixed traffic environment in Melbourne \cite{VicroadsReport2013}. }


\section{Simulation Results}

\subsection{Tram Performance}

\begin{figure}[!h]
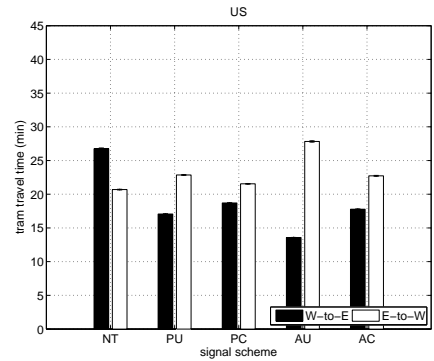
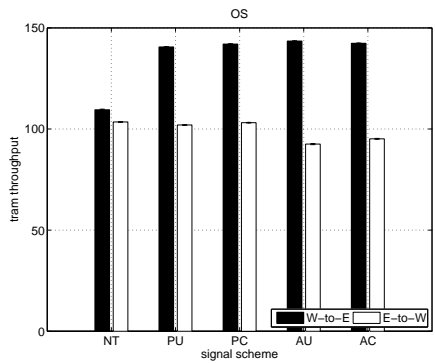
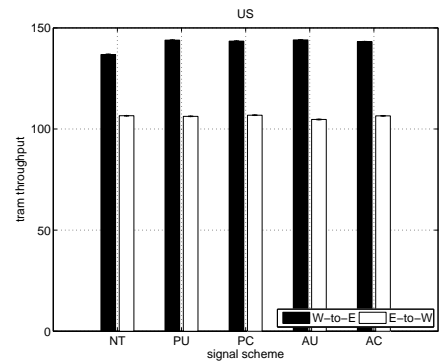
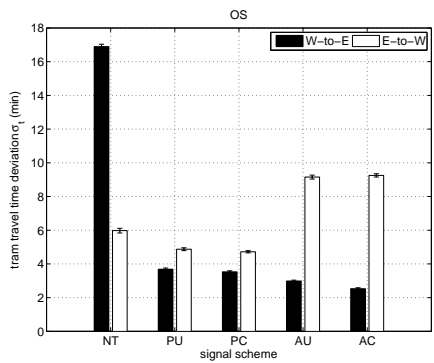
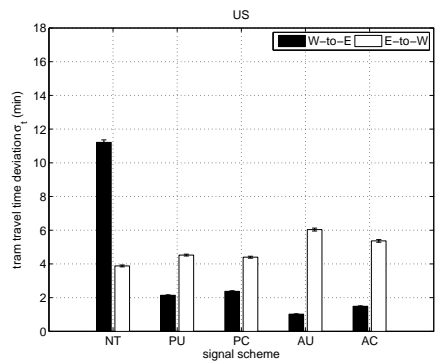

\centering
\subfigure[Mean tram travel time (OS)]{\tramageOS} \subfigure[Mean tram travel time (US)]{\tramageUS}
\subfigure[Mean tram throughput for all 4 routes (OS)]{\tramthroughputOS} \subfigure[Mean tram throughput for all 4 routes (US)]{\tramthroughputUS}
\subfigure[Tram travel time variability (OS)]{\tramdevOS} \subfigure[Tram travel time variability (US)]{\tramdevUS}
\vspace{-0.3cm}
\caption{Tram performance. Error bars corresponding to one standard deviation are shown {but are usually too small to observe}.}
\label{fig:tramPerformance}
\end{figure}

Fig.~\ref{fig:tramPerformance} compares the tram performance {for various signal schemes. As expected, tram priority reduces the average tram travel time in the eastbound direction, when compared to the no priority system \nt}, and the improvement is more significant under the OS scenario than US. It is also unsurprising that the \au\ scheme produces the largest improvements, {saving about 56\% and 48\% eastbound travel times under OS and US respectively}. \au\ achieves the goal that trams traveling in the peak (priority) direction are rarely delayed, even in congested conditions. However, it results in significant delays for westbound trams. The average travel times under the \tu\ system lie in-between the results of \nt\ and \au. The performance of eastbound trams is improved whilst westbound trams does not suffer significant delays. The conditional systems \tc\ and \ac, compared to their unconditional versions \tu\ and \au\ separately, result in slightly longer eastbound travel times, but have less impact on the westbound direction. 

In terms of throughputs, the priority systems produce essentially the same results. \nt\ results in a marginally lower eastbound throughput than other systems in the US scenario. This discrepancy increases significantly in the OS case.

Bus priority only produces significant delay savings at high levels of saturation \cite{JepsonFerreira00}. By contrast, tram priority achieves great savings in both unsaturated and over-saturated scenarios. This is because trams are more likely to be affected by cars, especially right-turning ones. Since trams cannot change lane, a single right-turning car, which has to give way to opposite traffic on the mixed traffic lane during phase $\ra$ (or $\rb$) could block the tram for an entire phase and cause a significant delay. Therefore, tram priority plays a significant role in determining tram performance. For \nt, the large delay in eastbound tram travel times is due to insufficient running time for phase $\re$ (or $\ret$). Analogously, the \au\ scheme provides the largest delay in westbound travel times, because it does not allocate sufficient time for phase $\rf$. The partial priority schemes balance the demands for $\re$ (or $\ret$) and $\rf$ and give acceptable results for both directions. 

In addition to improve travel times and throughputs, tram priority significantly reduces eastbound travel time variability. Similar to the travel time results, the absolute priority systems provide the best result in the eastbound direction and the worst result in the westbound direction. 
In the OS scenario, the two partial priority schemes outperform \nt\ in both directions.

\subsection{Private Vehicle Performance}

\begin{figure}[!h]
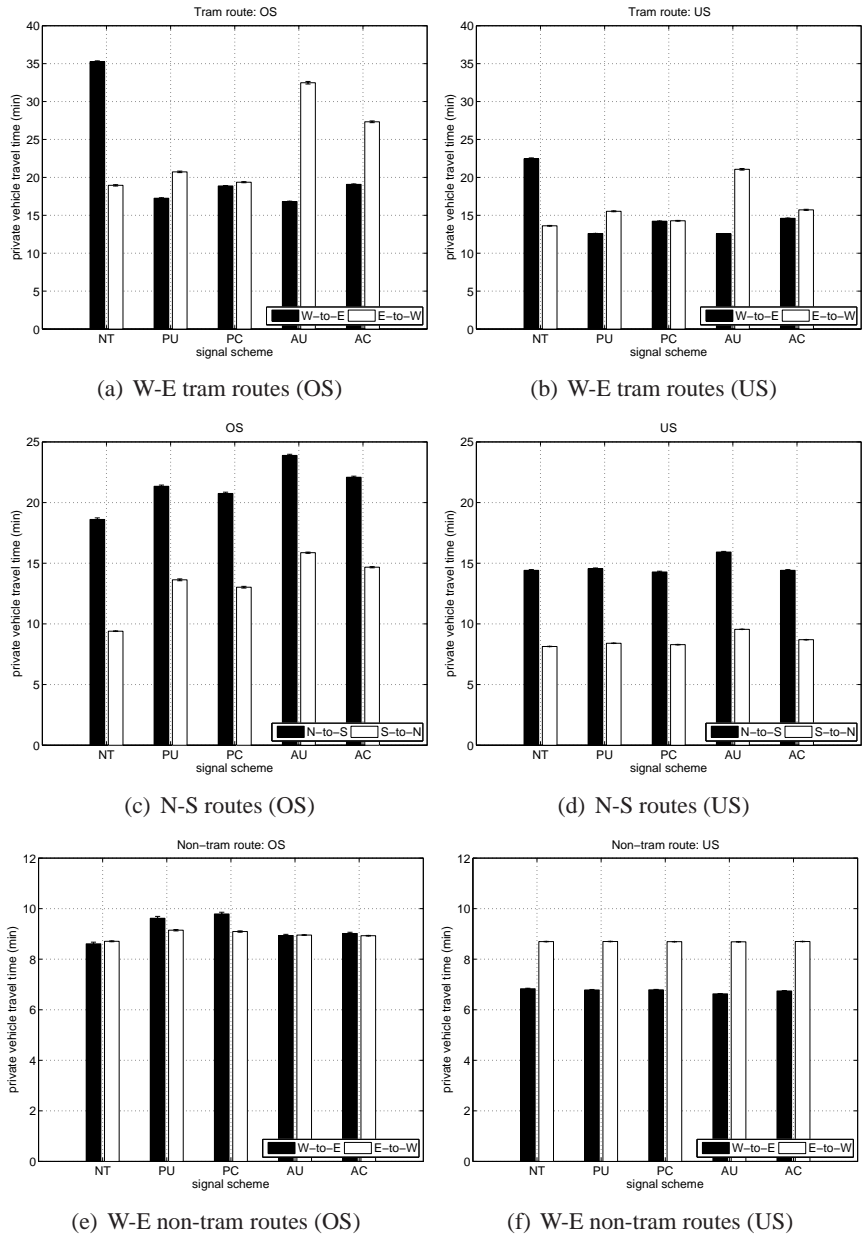

\centering
\subfigure[W-E tram routes (OS)]{\carageWEOST} \subfigure[W-E tram routes (US)]{\carageWEUST}
\subfigure[N-S routes (OS)]{\carageNSOS} \subfigure[N-S routes (US)]{\carageNSUS}
\subfigure[W-E non-tram routes (OS)]{\carageWEOSN} \subfigure[W-E non-tram routes (US)]{\carageWEUSN}
\vspace{-0.3cm}\caption{Mean car travel times. Error bars corresponding to one standard deviation are shown {but are usually too small to observe}.}
\label{fig:carAge}
\end{figure}

Fig.~\ref{fig:carAge} shows the mean travel time of cars traveling along different approaches. For the west-east direction, we separate cars that have traveled along non-tram routes from those along tram routes. We remark that a car is considered to travel along a tram route only if it traverses the whole route without turning into other links. Similar definitions are used for cars traveling in different directions. 

The performance of cars traveling along tram routes is quite similar to that of trams. Although the inflow rate of cars on tram routes is much less than that on non-tram routes, the car travel time along tram routes is much longer, which is partly due to trams and partly due to right-turning cars at nodes. Right-turning cars result in capacity drops at tram nodes, since there are no exclusive right-turning lanes on tram routes and such vehicles are required to give way to opposing traffic during phases $\ra$ and $\rb$ and so hinder other straight-going vehicles behind them. 

As expected, when tram priority process is active, regardless of the scheme used, both southbound and northbound travel times increase. The higher the priority imposed, the more the north-south traffic gets penalized. Even though the \au\ and \ac\ schemes penalize all three non-priority directions, they penalize the north-south traffic more than \tu\ and \tc\ do. The impact on the north-south traffic is rather negligible when the network is unsaturated however. This is because \scats\ uses adaptive split plans; when the congestion in the north-south direction grows as a result of giving priority to trams, it adjusts the split plan and assigns more split time to the north-south phase after priority process is complete. Nevertheless, this adaptivity becomes less effective when the tram volume is high and/or the tram route is over-saturated. 

Interestingly, we observe from Fig.~\ref{fig:carAge}(e) that in the OS regime tram priority can penalize the traffic in parallel non-tram routes. Perhaps surprisingly, the penalty generated by \tu\ and \tc\ is larger than that by \au\ and \ac. This arises because absolute tram priority results in larger decreases in both the north-south flow and the amount of traffic turning into the east-west direction, which therefore induces an effective \em{gating} of the west-east non-tram routes. 
This type of unexpected non-local behavior illustrates the importance of studying the response of the network as a whole, rather than just focusing on the route on which priority is being imposed.
Finally, we note that the reason the mean travel time along the eastbound non-tram routes is always less than that along the southbound routes, both of them are peak directions, is simply a consequence of signal linking being applied in the eastbound direction.



\subsection{Person Performance}

One aim of traffic management for road authorities is to move as many people as possible in each lane in order to maximize the use of road. Compared to private vehicles, trams have higher occupancy levels, which is a key motivation of tram priority. In this subsection we address a quantitative question, which is what tram occupancy is to justify various tram priority levels in different scenarios. Specifically, We pinned the car occupancy $o_c$ and the ratio $o_{t,p}/o_{t,n}$ of tram occupancy in the peak and counter-peak directions, and studied people travel time and throughput as a measure of network performance with various $o_{t,p}$.

\begin{figure}[!h]
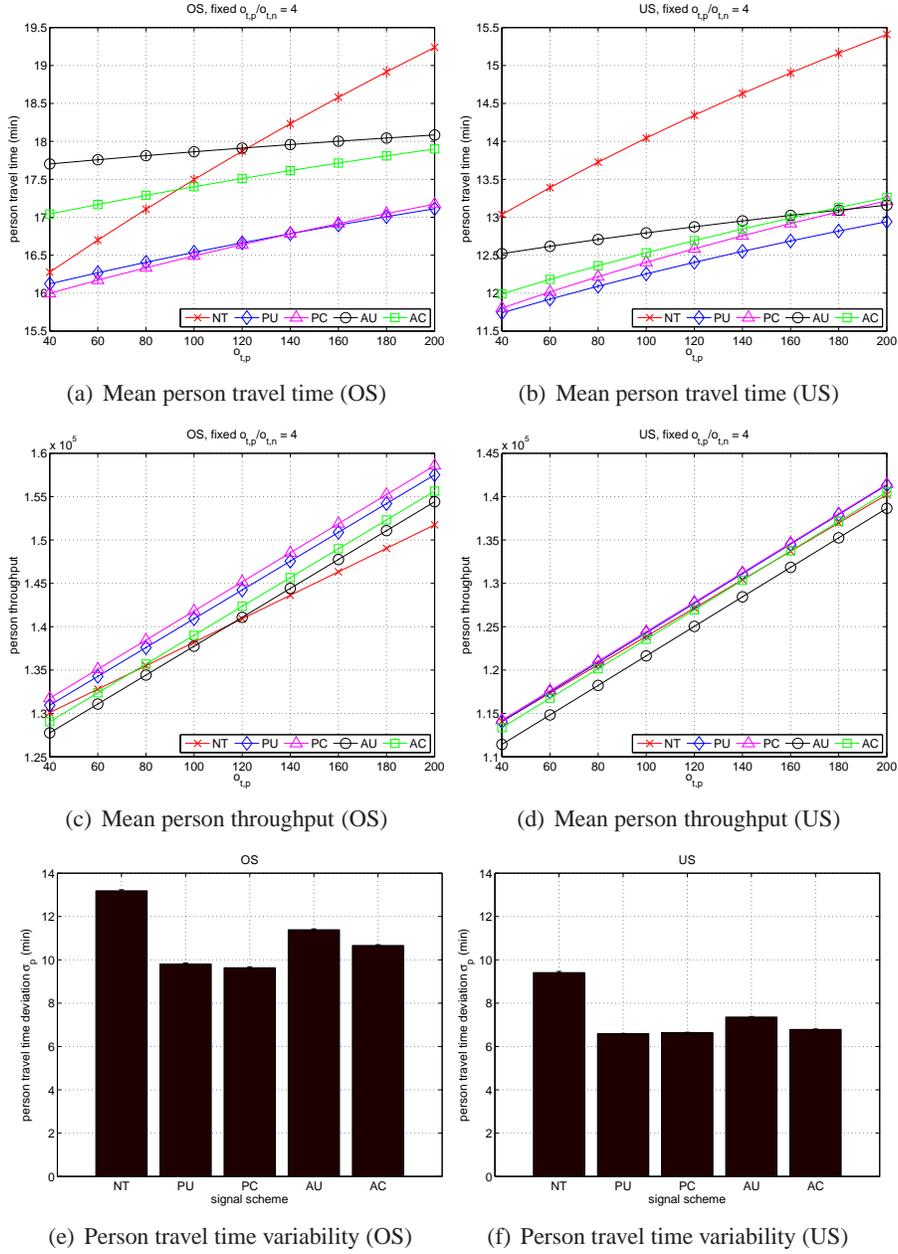

\centering
\subfigure[Mean person travel time (OS)]{\peopleageOSALL} \subfigure[Mean person travel time (US)]{\peopleageUSALL}
\subfigure[Mean person throughput (OS)]{\peoplethroughputOSALL} \subfigure[Mean person throughput (US)]{\peoplethroughputUSALL}
\subfigure[Person travel time variability (OS)]{\peopledevOSALL} \subfigure[Person travel time variability (US)]{\peopledevUSALL}
\vspace{-0.3cm}\caption{Person performance. (a)-(d): $o_{t,p}/o_{t,n}=4$ and $o_{t,p} = 40,60,\dots, 200$. (e) and (f): $o_{t,p}=80$ and $o_{t,n}=20$. Error bars corresponding to one standard deviation are shown {but are usually smaller than the symbol size of the data point}.}
\label{fig:ppPerformanceOverall}
\end{figure}

Figs.~\ref{fig:ppPerformanceOverall}(a)-(d) give the average person travel time and throughput of the whole network as $o_{t,p}$ varies. The differences in person throughputs between the various systems are very limited. In short, \au\ provides the worst result in the US case and for $o_{t,p}\geq 120$ in the OS case, whereas \tc\ always provides the best result in both cases. Next we focus on the result of person travel times.

For the unsaturated network, regardless of the tram occupancy, all the priority schemes result in reduced person travel times. The improvement is more pronounced when $o_{t,p}$ is larger, as expected. \au\ outperforms \ac\ and \tc\ when $o_{t,p}\geq 180$. This implies that from the perspective of individual travelers, implementing tram priority processes in this case improves tram performance to an extent which overweighs the negative impact on other traffic, and overall it has a positive effect on the network. We expect the crossing point determining when the absolute priority becomes the optimal scheme should move to lower occupancy as we decrease the congestion. This is because for low congestion the signals are adaptive enough to cope with the penalties caused by tram priority. 

In the saturated scenario, \tu\ and \tc\ obtain the smallest travel times for all reasonable values of tram occupancy. The travel time curve for \nt\ intersects with \ac\ and \au\ at $o_{t,p}=100$ and $o_{t,p}=120$ separately. This implies that although the absolute tram priority schemes bring relatively large penalties to other road users, compared to no priority scheme, they provide better overall network efficiency in terms of person travel times when tram occupancy is sufficiently high.

With respect to the relation between conditional and unconditional schemes, we see that the variability for \tu\ and \tc\ is almost identical in both traffic conditions, yet travel times and throughputs for \tc\ are marginally better. Combining this with the results in Figs.~\ref{fig:tramPerformance} and \ref{fig:carAge}, it appears that \tc\ obtains most of the benefit obtained by \tu\ but with a slightly lower penalty on the network, and therefore arguably produces an overall better result. By contrast, compared with \au, \ac\ unambiguously performs better in terms of person travel times, variability and throughputs, and should be therefore preferred.

\section{Conclusion}

We have utilized a multimodal traffic model to study a variety of tram priority schemes in a mixed traffic environment on a square lattice network. In particular we have studied the adaptive traffic signal system \scats\ with a number of tram priority scenarios, using a morning-peak traffic profile and two orthogonal peak directions. We have considered two scenarios with low and high levels of saturation.

Tram priority is an effective strategy to improve tram performance in terms of both travel time and variability. Regardless of the traffic condition, the absolute tram priority results in the best tram service in the priority direction at the expense of delaying other traffic in the non-priority directions. 
At a lower level of saturation, the impact of tram priority on the orthogonal direction is almost negligible, which can be explained by the adaptivity of the traffic signal system. In this case, with high tram occupancy the absolute tram priority might be justified.
As the network becomes more saturated however, other road users suffer more from the disruptions caused by tram priority processes, especially the absolute tram priority.

With respect to the overall person performance, the partial priority gives the best result. In general the partial priority should be recommended. The savings for priority-direction traffic derived from the absolute priority is negated by the costs imposed on opposing traffic, unless trams have extremely high occupancy. 
For either the absolute priority or the partial priority, the conditional version achieves almost the same level of improvement of service as the unconditional version but with reduced impact on other traffic. Therefore, the partial conditional priority system appears worth trialling. In the case that the absolute tram priority is necessary, e.g. in order to keep tram service on time regardless of the traffic condition, the absolute conditional priority should be implemented, rather than the absolute unconditional. 

The analysis of tram priority presented in this paper is just a first attempt at using the multimodal traffic simulation model on large-scale networks. Future work will extend the study of the tram priority to two directions: both peak and counter-peak for all the priority schemes. This is challenging since counter-peak-direction tram priority may disadvantage peak-direction trams. We also intend to study advanced priority schemes for intersecting tram routes.
{In addition we have made some assumptions to calibrate the model, including constant linking offset, constant expected tram travel speed and fixed frequency of phases $\re$ and $\rf$. In practice, those could depend on traffic conditions. Future work will consider the impact of those parameters.}

\section{Acknowledgments}
We gratefully acknowledge the financial support of the Roads Corporation of Victoria (VicRoads), and we thank VicRoads staff, in particular Adrian George, Andrew Wall, Anthony Fitts, Hoan Ngo and Chris Eer for numerous valuable discussions. This work was supported under the Australian Research Council's Linkage Projects funding scheme (project number LP120100258), and T.G. is the recipient of an Australian Research Council Future Fellowship (project number FT100100494). This research was undertaken with the assistance of resources provided at the NCI National Computational Merit Allocation Scheme supported by the Australian Government. We also greatly acknowledge access to the computational facilities provided by the Monash Sun Grid.

\bibliographystyle{dcu}
\bibliography{TranScience}

\appendix
{
\section{SCATS -- Cycle Length Decision}

Every time a master node is about to restart its cycle, the cycle length is adjusted adaptively based on recent measured values of the \ds.
In our model of \scats, the \ds\ of in-lane $\lambda$ and phase $\mp$ is defined to be
\begin{align}
{\ds}_{\lambda,\mp} &= \frac{1}{\ms_\mp}\left[\ms_\mp-{\sum_{t=1}^{\ms_\mp} (1-{o_\lambda(t)})}+ {N}_\lambda{\sum_{t=1}^{\ms_\mp} F_\lambda(t)}\right],
\label{equ:DS}
\end{align}
where $o_\lambda(t)$ and $F_\lambda(t)$ are the stop-line occupancy and flow through the intersection of lane $\lambda$ at time $t$ respectively, and $\ms_\mp$ is the split time of $\mp$.
The quantity $N_\lambda$ denotes a fixed benchmark of the time required to traverse the gap between vehicles at maximum flow $F_{\max}$. 

The \ds\ of a master node, $\rm m$, is given by
\begin{align}
\ds_{\rm m}&=\ds_{\mp^*}=\max_{\lambda^*}\ds_{\lambda^*,\mp^*},
\end{align}
where the maximum is taken over all in-lanes in the linked direction during linked phase $\mp^*$.

For a non-subsystem node, $\rm n$, 
the \ds\ is defined by the maximum \ds\ over all in-lanes and phases,
\begin{align}
\ds_{\rm n}&=\max_{\mp}\ds_{\mp}=\max_{\mp}\max_{\lambda} \ds_{\lambda,\mp}.
\end{align}
We remark that an in-lane is excluded from the computation of \ds\ if it belongs to any of the dashed paths shown in Fig.~\ref{fig:phases}, to avoid pathological \ds\ values induced by turning vehicles.

At a given time, the weighted \ds\ of the previous 3 cycles is defined to be
\begin{align}
{\rm wDS} &= 45\%\ds_0+33\%\ds_{-1}+22\%\ds_{-2},
\end{align}
where $\ds_{1-i}$ is the \ds\ of the $i$th last cycle. 


The strategy for adapting the cycle length $\mc$ based on $V=\ds \times F_{\max}$ and $\text{w}\ds$, of a master (or non-subsystem) node is as follows.

\begin{myalg}{\scats\ cycle length decision}\label{alg:scats}

\smallskip
\begin{tabular}{p{11.5cm}}
\hline
{\bf Case 1: \quad if }  $\mc=\mc_{\tt MIN}$  $\& \,V > 0.4$, {\bf then} $\mc = \mc_{\tt STOPPER}$\\
{\bf Case 2: \quad if }  $\mc=\mc_ {\tt STOPPER}$  $\& \,V < 0.2 $, {\bf then} $\mc =\mc_{\tt MIN}$\\
{\bf Case 3: \quad if }  ${\rm w}\ds> 0.95$, {\bf then} $\mc=\min\{\mc+\mc_{\tt STEP}, \mc_{\tt MAX}\}$ \\
{\bf Case 4: \quad if } ${\rm w}\ds <0.85$, {\bf then} $\mc=\max\{\mc-\mc_{\tt STEP}, \mc_{\tt STOPPER}\}$ \\
{\bf Otherwise: } \quad$\mc$ remains unchanged.\\
{\rm $\mc_{\tt MIN}=48$sec; $\mc_{\tt MAX}=134$sec; $\mc_{\tt STOPPER}=68$sec; $\mc_{\tt STEP}=6$sec.}\\
\hline
\end{tabular}
\end{myalg}

The $\mc_{\tt STOPPER}$ is included to allow a steep increase in the cycle length due to a sudden increase in traffic volume, when the cycle length is at its minimum.
}

\end{document}